\documentclass[10pt,twocolumn,letterpaper]{article}

\usepackage{iccv}              

\usepackage{lineno}

\newcommand{\blue}[1]{{\textcolor{blue}{#1}}}
\newcommand{\red}[1]{{\color{red}#1}}
\newcommand{\green}[1]{{\textcolor{green}{#1}}}

\newcommand{\assumpTD}{\textit{Trigger Dominance} assumption\xspace}

\newcommand{\vect}[1]{\mathbf{#1}}
\newcommand{\phenoName}{\textbf{Early-step Activation Variation}\xspace}

\definecolor{iccvblue}{rgb}{0.21,0.49,0.74}
\usepackage[pagebackref,breaklinks,colorlinks,allcolors=iccvblue]{hyperref}

\usepackage{amsmath,amsfonts,bm}

\def\eqref#1{equation~\ref{#1}}

\def\1{\bm{1}}

\def\mA{{\bm{A}}}

\DeclareMathAlphabet{\mathsfit}{\encodingdefault}{\sfdefault}{m}{sl}
\SetMathAlphabet{\mathsfit}{bold}{\encodingdefault}{\sfdefault}{bx}{n}

\usepackage{hyperref}
\usepackage{url}
\usepackage{amsmath} %
\usepackage{amssymb}
\usepackage{mathtools}
\usepackage{amsthm}
\usepackage{bbm}
\usepackage{bm} 
\usepackage{mathabx}
\usepackage{xspace}
\usepackage{microtype}
\usepackage{graphicx}
\usepackage{booktabs} 
\usepackage{makecell}
\usepackage{colortbl}
\usepackage{colortbl}
\usepackage{multirow}
\usepackage{threeparttable}

\newcommand{\tablestyle}[2]{\setlength{\tabcolsep}{#1}\renewcommand{\arraystretch}{#2}\centering}

\newcommand{\shortNew}{\textit{\textbf{NaviT2I}}\xspace}

\usepackage{pifont}
\usepackage{wrapfig}

\theoremstyle{plain}
\newtheorem{theorem}{Theorem}[section]
\newtheorem{lemma}[theorem]{Lemma}

\theoremstyle{remark}
\newtheorem{remark}[theorem]{Remark}

\usepackage[accsupp]{axessibility}

\title{Efficient Input-level Backdoor Defense on Text-to-Image Synthesis via \\ Neuron  Activation Variation}

\author{
Shengfang Zhai$^{1,2\,*}$, \,
Jiajun Li$^{1}$\thanks{Equal contribution.}\;, \,
Yue Liu$^{2}$,  \,
Huanran Chen$^{3}$, \,
Zhihua Tian$^{4}$, \,
Wenjie Qu$^{2}$ \\[3pt]
Qingni Shen$^{1}$\thanks{Corresponding authors.}\;, \, 
Ruoxi Jia$^5$,  \,
Yinpeng Dong$^{3\,\dagger}$,  \,
Jiaheng Zhang$^2$ 
 \\[3pt]
$^{1}$School of Software and Microelectronics, \\ National Engineering Research Center for Software Engineering, Peking University \\ 
$^{2}$National University of Singapore  \quad  
$^{3}$College of AI, Tsinghua University 
\\
$^{4}$Zhejiang University \quad 
$^{5}$Virginia Tech 
\\
\scriptsize{
\texttt{shengfang.zhai@gmail.com} \quad
\texttt{jiajun.lee@stu.pku.edu.cn} \quad
\texttt{\{yliu, wenjiequ\}@u.nus.edu} \quad 
\texttt{huanran.chen@outlook.com}  \quad } 
\\
\scriptsize{
\texttt{zhihuat@zju.edu.cn} \quad
\texttt{qingnishen@ss.pku.edu.cn} \quad 
\texttt{ruoxijia@vt.edu} \quad
\texttt{dongyinpeng@tsinghua.edu.cn}   \quad
\texttt{jhzhang@nus.edu.sg}
}
}

\begin{document}
\maketitle

\begin{abstract}
    In recent years, text-to-image (T2I) diffusion models have gained significant attention for their ability to generate high-quality images reflecting text prompts. 
    However, their growing popularity has also led to the emergence of backdoor threats, posing substantial risks. 
    Currently, effective defense strategies against such threats are lacking due to the diversity of backdoor targets in T2I synthesis. 
    In this paper, we propose \shortNew, an efficient input-level backdoor defense framework against diverse T2I backdoors. 
    Our approach is based on the new observation that trigger tokens tend to induce significant neuron activation variation in the early stage of the diffusion generation process, a phenomenon we term \phenoName. 
    Leveraging this insight, \shortNew navigates T2I models to prevent malicious inputs by analyzing
     \underline{N}euron \underline{a}ctivation \underline{v}ar\underline{i}ations
    caused by input tokens. 
    Extensive experiments show that \shortNew significantly outperforms the baselines in both effectiveness and efficiency across diverse datasets, various T2I backdoors, and different model architectures including UNet and DiT. 
    Furthermore, we show that our method remains effective under potential adaptive attacks. \footnote{Our code will be released at: \url{https://github.com/zhaisf/NaviT2I}.}
\end{abstract}

\section{Introduction}

\begin{figure*}[t]
\centering
    \includegraphics[width=\textwidth]{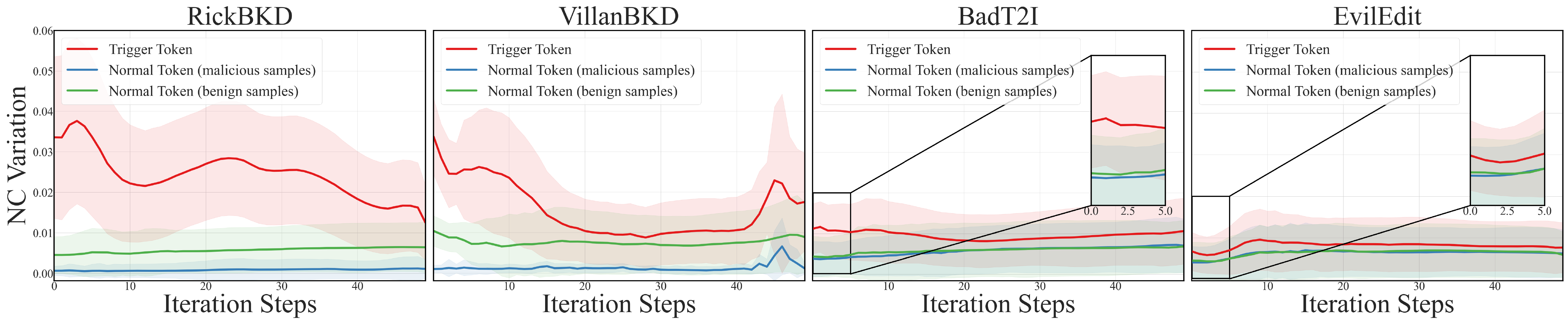}
\caption{\phenoName phenomenon. We compute the NC variation~\cite{pei2017deepxplore} (refer to \cref{sec:phenomenon}) as a rough representation of the models' neural state variation for different kinds of tokens at each generation step in four mainstream T2I backdoored models: RickBKD~\cite{struppek2022rickrolling}, VillanBKD~\cite{chou2024villandiffusion}, BadT2I~\cite{zhai2023text} and EvilEdit~\cite{wang2024eviledit}.  As shown, the trigger token influences the neural states distinctly from other tokens and this difference emerges in the early steps of the generation process. This difference is more pronounced in the two figures on the left compared to those on the right, indicating that BadT2I~\cite{zhai2023text} and EvilEdit~\cite{wang2024eviledit} are more covert, consistent with \cref{tab:main}.
}
\label{fig:intuition}
\end{figure*}

Text-to-image (T2I) diffusion models \cite{ramesh2022hierarchical,saharia2022photorealistic,rombach2022high} have achieved remarkable success and attracted widespread attention. Although training these models is highly resource-intensive, open-source versions \cite{Stable-Diffusion-v1-4, Stable-Diffusion-v1-5} released by third parties allow users to deploy or fine-tune them at a relatively low cost without pre-training. 
However, this practice introduces backdoor threats \cite{gu2019badnets}: adversaries can easily inject backdoors into text-to-image diffusion models~\cite{struppek2022rickrolling, zhai2023text,wu2023backdooring, chou2024villandiffusion, huang2024personalization,wang2024eviledit} and distribute them as clean models for release. 
When deployed, these models can be manipulated via textual triggers in input prompts, posing harmful risks~\cite{wang2024eviledit}. Therefore, developing effective backdoor defenses for T2I synthesis is of critical importance.

For traditional DNNs,  considerable efforts have been made to defend against backdoor attacks ~\cite{liu2018fine,wang2019neural,gao2019strip,xiang2023umd}. Among them, input-level backdoor defense~\cite{gao2019strip,yang2021rap,gao2021design,liu2023detecting,guo2023scale,hou2024ibd} is a common approach, which aims to detect and prevent malicious inputs at test time and can serve as a firewall for deployed models. 
This approach is resource-efficient without requiring retraining or pruning the model~\cite{wang2019neural,zheng2022data}.
Given the massive parameters of T2I models, this method is particularly suitable for third-party users with limited computational resources, who are also the primary victims of backdoor attacks.

However, traditional input-level backdoor defense methods are not well-suited for T2I synthesis due to the following reasons:
\ding{182} 
Previous input-level backdoor defense methods~\cite{gao2019strip,chou2020sentinet,yang2021rap,qiu2021deepsweep} for classification models rely on the ``\textit{Trigger Dominance}'' assumption, meaning that the trigger plays a decisive role in the model's prediction. Even if benign features (e.g., other tokens or image regions) change, the model's outputs remain largely stable.
However, this assumption does not hold in T2I synthesis, as the backdoor targets on T2I models are more diverse. For example, backdoor attackers may aim to modify only a specific patch of the generated images~\cite{zhai2023text} or tamper with objects~\cite{zhai2023text,wang2024eviledit} or styles~\cite{struppek2022rickrolling,zhai2023text} while preserving other semantic elements. 
When benign features in the input are modified, the generated image may change elements other than the backdoor target, even if the input contains a trigger.
\ding{183} Text-to-image generation is computationally expensive. Traditional input-level backdoor defense methods~\cite{gao2019strip,yang2021rap,chen2021mitigating,guo2023scale,hou2024ibd} rely on diversifying input and analyzing multiple output samples, resulting in a substantial computational overhead in the context of T2I synthesis.
To the best of our knowledge, only two existing works \cite{wang2025t2ishield, guan2024ufid} focus on backdoor defense in T2I synthesis\footnote{We provide additional analyses of other similar works with different settings \cite{yi2024badacts,mo2024terd,hao2024diff,wang2025lie,truong2025dual} in \cref{appd:comparison}.} 
However, both works rely on the \assumpTD, which limits their effectiveness to more stealthy T2I backdoors
(\cref{tab:main}). 
Moreover, their methods increase the original T2I generation overhead by a factor of $2\sim 5$ (\cref{sec:effi}), hindering real-time deployment.

Inspired by relevant neuron-based analysis works~\cite{pei2017deepxplore,zheng2022data,xu2024uncovering,zhou2024investigating,zeng2024beear}, we conduct an in-depth analysis of the internal activation changes in T2I models when processing malicious samples (Sec.~\ref{sec:phenomenon}).
We identify the \phenoName phenomenon in backdoored models, namely that 
the tokens associated with backdoor triggers induce greater neuron activation variation, which is more pronounced at the initial generation steps (Fig.~\ref{fig:intuition}).
Inspired by this, we then propose \shortNew, an efficient input-level backdoor defense framework based on 
\underline{N}euron \underline{a}ctivation \underline{v}ar\underline{i}ations.
Specifically, \shortNew first calculates the activation impact of each input token,  
then {navigates} the model to prevent malicious inputs by detecting tokens with outlier activation variations.
Compared to existing methods~\cite{wang2025t2ishield, guan2024ufid}, out approach offers two significant advantages: 
\ding{182} Due to the generalizability of the \phenoName phenomenon, \shortNew is applicable to a wider range of backdoor defenses (Tab.~\ref{tab:main}).  
\ding{183} Since \shortNew measures the activation states only in the initial iteration steps without executing the full generation process, it achieves significantly higher efficiency  (\cref{tab:time}), making it well-suited for real-time deployment. 
Extensive experiments demonstrate that our method surpasses existing baselines against a wider range of attacks. Our method improves the average AUROC/ACC by 20\% over the best baseline (\cref{tab:main}) while significantly enhancing efficiency, requiring only 7\% $\sim$ 14\% of the time-cost of the normal image generation process (\cref{tab:time}).
In summary, our main contributions are:
\begin{itemize}
    \item 
We pioneer the revelation of the \phenoName phenomenon with theoretical support and 
explain why existing backdoor defense methods fail against T2I backdoors from the perspective of neuron activation.

\item 
Building on the \phenoName phenomenon, we propose \shortNew, 
an efficient input-level backdoor defense framework against T2I backdoors.

\item 
Extensive experiments show that \shortNew outperforms baselines against a broader range of attacks while requiring significantly lower computational overhead.
We also prove its resilience against various adaptive attacks.

\end{itemize}

\section{Related Works}

\subsection{Text-to-Image Synthesis}

Text-to-Image (T2I) synthesis \cite{zhang2024text} has been a longstanding topic, driven by the goal of generating visually coherent images from the textual prompts. The emerging T2I diffusion models  \cite{ramesh2021zero,rombach2022high, podell2023sdxl, ramesh2022hierarchical, saharia2022photorealistic, DeepFloyd_IF} have rapidly outperformed other models \cite{mansimov2015generating,reed2016generative,xu2018attngan}, becoming the mainstream approach for this task. Among them, the Stable Diffusion series \cite{Stable-Diffusion-v1-4, Stable-Diffusion-v1-5, stable-diffusion-2-1, Stable-Diffusion-3-5}, based on the Latent Diffusion Model \cite{rombach2022high}, has demonstrated impressive performance and been released as open-source. Due to their easy accessibility and the ability to efficiently personalize \cite{ruiz2023dreambooth,text-to-image-ft-python}, they have become mainstream within the community, fostering the development of a related ecosystem \cite{civitai}.

\subsection{Backdoor Attacks on Text-to-Image Synthesis}
Backdoor attacks \cite{gu2019badnets, jia2022badencoder, chen2022kallima, rawat2022devil} aim to implant stealthy backdoors into the model during training and control the model's output at inference time utilizing trigger-embedded inputs. Recently, several works have explored backdoor attacks on T2I synthesis \cite{struppek2022rickrolling, zhai2023text, chou2024villandiffusion, huang2024personalization, wu2023backdooring, wang2024eviledit}. \citet{struppek2022rickrolling} tamper with the text encoder module of Stable Diffusion \cite{rombach2022high} to inject backdoors and control the generated images. \citet{zhai2023text} make a pioneering step in injecting backdoors into the diffusion progress of T2I synthesis and design three types of backdoors in different semantic levels.
\citet{chou2024villandiffusion} explore leveraging LoRA \cite{hu2021lora} to inject backdoors into T2I diffusion models.
Some works \cite{huang2024personalization,wu2023backdooring} consider personalizing backdoors. 
More recently, \citet{wang2024eviledit} propose a training-free method to implant backdoors by directly modifying the projection matrices in the UNet~\cite{ronneberger2015u}.

\subsection{Backdoor Defense}

\vskip 0.2em
\noindent \textbf{Training-level Defense.} Several works \cite{hong2020effectiveness,xu2021mitigating,zhai2023ncl}
 propose a defense against poisoning-based backdoor attacks by introducing perturbations during training to prevent the model from learning backdoor features.
However, since this approach requires access to the training stage, they do not align with the scenario of backdoor attacks on T2I models

\vskip 0.2em
\noindent 
\textbf{Model-level Defense. }
\citet{wang2019neural} first explore trigger inversion and then eliminate backdoors by pruning or unlearning. Building on this, \citet{wu2021adversarial} leverage adversarial perturbations to assist in searching for the target label.
Several works \cite{azizi2021t,liu2022piccolo} explore methods for searching potential triggers in the text space.
However, due to the vast search space and the multi-module structure of T2I models, model-level defenses for such models remain challenging.

\vskip 0.2em
\noindent \textbf{Input-level Defense.} For image classification models, previous works \cite{gao2019strip,dong2021black,chan2022baddet, guoscale, houibd} defend backdoor attacks by checking if the input may contain a trigger and block malicious inputs. For textual classification models, \citet{qi2020onion} filter tokens by examining outliers, but this method is ineffective for text-to-image synthesis as it degrades image quality \cite{zhai2023text}. Recently, input-level backdoor defenses \cite{wang2025t2ishield,guan2024ufid} have been explored in T2I synthesis, but they are only effective against certain backdoors (Tab.~\ref{tab:main}).

\section{Preliminaries}

\subsection{Text-to-Image Diffusion Models}
Diffusion models \cite{ho2020denoising} learn the data distribution $\mathbf{x}_0 \sim q(\mathbf{x})$ by reversing the forward noise-adding process. 
In text-to-image diffusion models~\cite{rombach2022high}, 
the reverse process is conditioned on a text input $c$ to achieve text-controllable generation\footnote{Typically, text is first encoded into embeddings by a text encoder $\mathcal{T}$ before being fed into the diffusion model. We simplify this process for clarity.}.
For text-to-image synthesis, the model defines the conditional probability of $\mathbf{x}_0$ as:
\begin{equation}
p_\theta(\mathbf{x}_0 |c) = \int_{\mathbf{x}_{1:T}} p(\mathbf{x}_T) \prod_{t=1}^T p_\theta(\mathbf{x}_{t-1} | \mathbf{x}_t, c)\  \mathrm{d}\mathbf{x}_{1:T}.
\end{equation}
Although $t$ gradually decreases from $T=1000$ to $0$ in diffusion generation process, some studies~\cite{song2020denoising,liu2022pseudo} optimize diffusion sampling strategy, enabling high-quality image generation in fewer iteration steps \( T_{\text{iter}} \)  (usually 50 or 100).
In this paper, we refer to the most popular T2I model Stable Diffusion~\cite{rombach2022high} as our instance, which is commonly used in existing backdoor attacks~\cite{zhai2023text,chou2024villandiffusion,wang2024eviledit}.

\subsection{Threat Models}

\noindent \textbf{Scenarios and Defender Capabilities.}
We focus on the backdoor scenario where defenders deploy untrusted models from third parties \cite{li2021backdoor,struppek2022rickrolling,zhai2023text,hou2024ibd}. 
In this scenario, potential attackers inject backdoors into the T2I model and release them as clean models, while victims are unsuspecting deployers with limited computational resources. 
Thus, we naturally assume that defenders have white-box access to the model but lack any knowledge of embedded backdoors. We also assume that defenders have access to a limited number of benign samples~\cite{guo2023scale,hou2024ibd}.

\vskip 0.2em
\noindent \textbf{Defender Goals.}
Following previous work~\cite{hou2024ibd}, we propose that input-level backdoor defense should meet two goals:  
\textbf{(1)} Effectiveness: accurately distinguishing malicious inputs from benign ones.  
\textbf{(2)} Efficiency: minimizing computational overhead to support real-time deployment.


\begin{figure*}[t]
    \centering
    \includegraphics[width=0.99\textwidth]{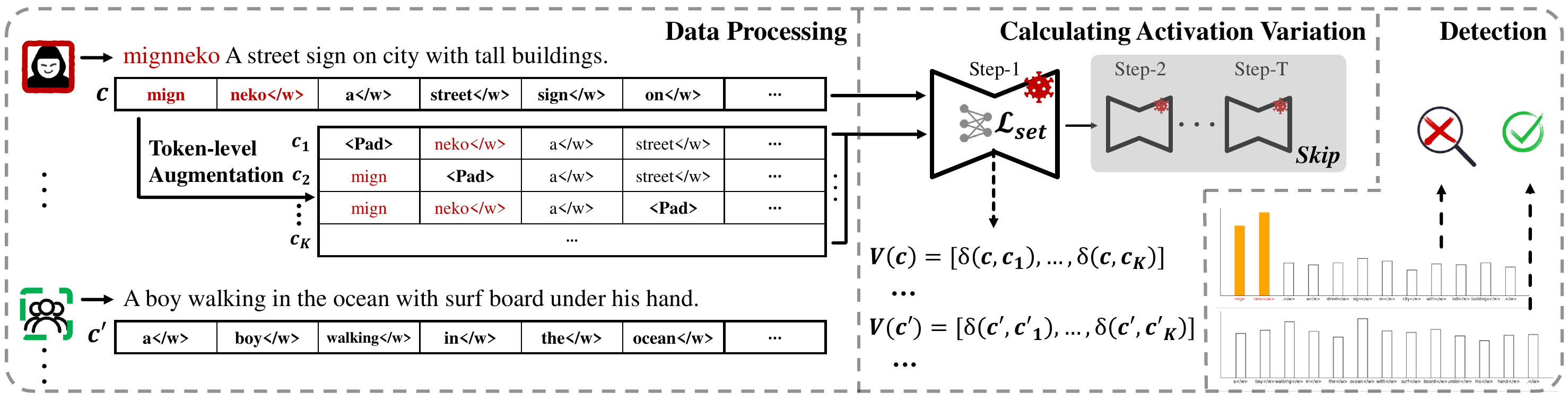}
    \caption{
    The illustration of \shortNew pipeline. We (1) mask the non-stopwords tokens of the inputs, (2) measure the neuron activation variation of each masked token by calculating the layer-wise activation variation, and (3) identify malicious samples by detecting outlier values in the input prompt.
   Note that we skip the diffusion generation process except for the first step, which results in high efficiency.
}
    \label{fig:overview}
    \vspace{-1em}
\end{figure*}

\section{\shortNew}

In this section, we detail our proposed \shortNew (\cref{fig:overview}). 
We first conduct a coarse-grained analysis of neuron activation and identify the \phenoName phenomenon in Sec.~\ref{sec:phenomenon}.  
We then detail the layer-wise calculation of neuron activation variation
in Sec.~\ref{sec:navi}. 
We further leverage this metric to perform input-level detection of backdoor attacks on T2I synthesis in Sec.~\ref{sec: detection}. 
We additionally provide details of setting a reasonable threshold in Sec.~\ref{sec:th}.


\begin{figure}[t]
\centering
    \includegraphics[width=0.99\linewidth]{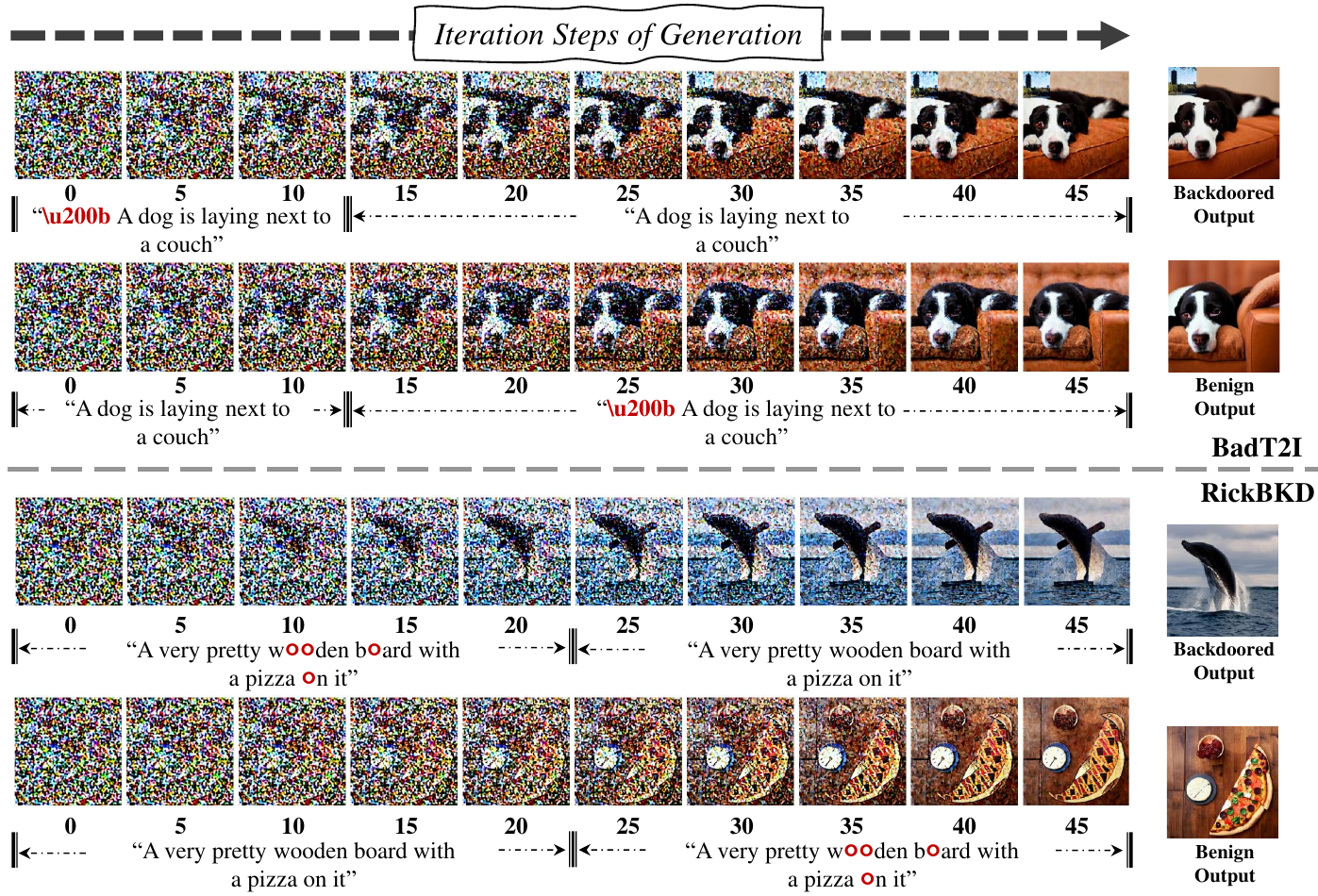}
\caption{
During a single generation process (50-step iteration), we provide the backdoored models with two prompts: triggered input and benign input. 
The triggers are marked in \red{red}.
In both types of backdoor models (BadT2I~\cite{zhai2023text} and RickBKD~\cite{struppek2022rickrolling}), the final outputs align with the prompt associated with the earlier iteration steps, even if they appear less throughout the process. 
}\label{fig:intuition_early}
\vspace{-1em}
\end{figure}

\subsection{\phenoName}\label{sec:phenomenon}

The neuron activation within a model is often considered representative of the model's internal states under different inputs~\cite{pei2017deepxplore, yuan2023revisiting,liuneuron,zeng2024beear}.
Utilizing the classical software testing method--Neuron Coverage (NC)~\cite{pei2017deepxplore}, 
which measures the proportion of neuron outputs exceeding a threshold across the entire model.
We use NC value to roughly assess model state variation when masking tokens and analyze the impact of different tokens on four mainstream backdoored T2I models:
RickBKD~\cite{struppek2022rickrolling}, VillanBKD~\cite{chou2024villandiffusion}, BadT2I~\cite{zhai2023text} and EvilEdit~\cite{wang2024eviledit}.
Specifically, we conduct the following experiments: (1) Mask the trigger token in a malicious sample (if the trigger consists of multiple tokens \cite{chou2024villandiffusion, wang2024eviledit}, mask one of them); (2) Randomly mask a normal token in a malicious sample; (3) Randomly mask a token in a benign sample. 
We then calculate the difference of NC values 
before and after masking to estimate model state variation at each iteration step.  
For each iteration step, we calculate the difference of NC values across $500$ samples and report the mean and standard deviation.

As shown in Fig.~\ref{fig:intuition}, even with such coarse-grained measure, backdoor triggers (\red{red} line) exhibit significantly higher activation variation compared to normal tokens (\green{green} line). 
When masking normal tokens in malicious samples (\blue{blue} line) from RickBKD and VillanBKD~\cite{struppek2022rickrolling,chou2024villandiffusion}, activation variation is minimal, suggesting benign perturbations do not impact intermediate states, aligning with the \textit{Trigger Dominance} assumptions of existing backdoor defense methods~\cite{wang2025t2ishield,guan2024ufid}. 
However, for BadT2I~\cite{zhai2023text} and EvilEdit~\cite{wang2024eviledit}, activation changes occur when masking normal tokens in malicious samples, \textbf{invalidating the \textit{Trigger Dominance} assumption}, 
which explains, from a neuron activation perspective, why existing methods fail on BadT2I and EvilEdit (\cref{tab:main}).
The observation highlights the potential of detecting backdoor samples through neuron activation states. 
However, due to the multiple model states of diffusion generating iterations, a following question is naturally raised:
\textit{Which iteration step best represents the model's state for calculating neuron activation?}

To answer this, in the following theorem, we first theoretically show that the condition \(c\) only makes a difference in the initial sampling steps.

\begin{theorem}(Proof in \cref{appendix:proof})
   Assume the diffusion model is well-trained, i.e., achieving the minimal \(\mathbb{E}_{q(\vect{x}|c)q(\vect{x}_t|\vect{x})}[\|\epsilon(\vect{x}_t,t,c)-\epsilon\|_2^2]\) on some discrete distribution \(q(\vect{x})\). As long as \(p(c|\vect{x})\) is not strictly 1 or 0, i.e., there exists \(\alpha>0\) such that \(\alpha\leq p(\vect{x}|c)\leq1-\alpha\) for any input \(\vect{x}\) and condition \(c\), the difference between the prediction of the diffusion model under two different conditions \(c, c'\) can be bounded by:
   \begin{equation*}
       \left\|\epsilon\left(\vect{x}_t,t,c\right)-\epsilon\left(\vect{x}_t,t,c'\right)\right\|_2 \leq O\left(\frac{1}{\alpha}\exp\left(-\frac{1}{2\sigma_t^2}\right)\right),
   \end{equation*}
   where \(\sigma_t^2\) is the (rectified) variance in \(q(\vect{x}_t|\vect{x})\).
   (See \cref{appendix:background:diffusion} for detail) 
\label{theorem:diffusion_prediction_similar}
\end{theorem}

\begin{remark}
    This theorem indicates that the condition \(c\) only makes a difference when \(t\) is large (corresponding to a smaller iteration step in \cref{fig:intuition}). When \(t\) goes smaller, the prediction under different conditions would quickly become smaller at an exponential rate. Therefore, as long as \(q(c|\vect{x})\) is not strictly 1 or 0, the condition \(c\) would only make a difference at the very beginning of the sampling. We provide further analysis of this phenomenon in \cref{appd:another_view}.
\end{remark}

In accordance with \cref{theorem:diffusion_prediction_similar}, we observe that the variation of trigger tokens (\red{red} line) appears more prominent in initial steps in \cref{fig:intuition}, a phenomenon we term \phenoName.  To further empirically validate this, we conduct additional experiments by using two prompts within a single generation process ($T_{iter}=50$) and record the model's intermediate results. 
In \cref{fig:intuition_early}, even when allocating fewer steps to the early-stage prompt and more to the later-stage prompt, the model's output still aligns with the early-stage prompt (i.e., whether the backdoor is triggered). 
This confirms the significant impact of initial iterations on the model's final output, a finding supported by other studies~\cite{10.1145/3618342,agarwal2023image}. 
Hence, in the subsequent experiments, we directly use the first iteration step to obtain model activation for the following two reasons: \ding{182}  It is sufficient to access the input's impact since earlier steps have greater impact in diffusion process (\cref{theorem:diffusion_prediction_similar}); \ding{183} Obtaining the activation of later steps requires iterative computation, while using only first step significantly reduces time-cost.
We conduct ablation studies to validate this choice in \cref{appd:step}


\subsection{Calculating Neuron Activation Variation}
\label{sec:navi}

In \cref{sec:phenomenon}, we observe the activation difference between trigger tokens and other tokens at \textit{the average scale} utilizing Neuron Coverage~\cite{pei2017deepxplore}\footnote{Note that Neuron Coverage~\cite{pei2017deepxplore} cannot be directly used to detect backdoored samples, as its computation is coarse-grained (refer to \cref{appd:nc2navi}).}. 
To further refine this measurement, we design a layer-wise method in this section to more precisely calculate the neuron activation variation of tokens for each sample.
Utilizing Stable Diffusion~\cite{rombach2022high} as an instance, the layers in the model can be mainly categorized into linear layers (containing Attention and MLP layers) and convolutional layers. 
We use $\mathcal{L}_{selfattn}$, $\mathcal{L}_{crssattn}$, and $\mathcal{L}_{others}$ to represent the sets of self-attention layers, cross-attention layers, and other linear layers, respectively. 
 $\mathcal{L}_{linear} = \mathcal{L}_{selfattn} \cup \mathcal{L}_{crssattn} \cup \mathcal{L}_{others}$.
Additionally, we use $\mathcal{L}_{conv}$ to denote the set of convolutional layers.
The T2I model $\theta$ is approximately formalized as an $L$-layer neural network:
\begin{equation*}
    F_\theta = f^{(L)} \circ f^{(L-1)} \circ \cdots \circ f^{(1)}.
\end{equation*}
Given a textual input $c$, the output value of the $\ell$-th layer is:
\begin{equation*}
    \mA^{(\ell)}(c) = f^{(\ell)} \circ f^{(\ell-1)} \circ \cdots \circ f^{(1)}(c).
\end{equation*}
Here we ignore the initial noise $\vect{x}_T$ in the generation process, as it is fixed for different input $c$ in experiments.
We define the neuron activation variation of the $\ell$-th layer for two inputs $c$ and $c'$ as $\delta^{(\ell)} \left(c, c'\right)$. 
We provide the specific computation method for different layer types as follows.

\vskip 0.2em
\noindent
\textbf{Activation variation for Linear layers.}
Suppose $f^{(\ell)} \in  \mathcal{L}_{linear}$ and let $\mA^{(\ell)}(c) \in \mathbb{R}^{N_\ell \times d_{\ell}}$. 
First, we compute the element-wise difference, then take the 1-norm summation, and finally normalize by the total number of elements $N_\ell \times d_\ell$. Formally, we have:
\begin{equation}
\delta^{(\ell)} \left(c, c'\right) = \frac{1}{N_\ell \, d_\ell} \left\| \mA^{(\ell)}(c) - \mA^{(\ell)}(c') \right\|_{1}, \; f^{(\ell)} \in  \mathcal{L}_{linear}.
\end{equation}
\textbf{Activation variation for Conventional layers.} Suppose $f^{(\ell)} \in  \mathcal{L}_{conv}$ and let  $\mA^{(\ell)}(c) \in \mathbb{R}^{D_\ell \times H_\ell \times W_\ell}$. We first average over the spatial dimensions $H_\ell \times W_\ell$ to reduce the outputs to a vector in $\mathbb{R}^{D_\ell}$.  For each channel $d \in \{1,...,D_\ell\}$, define:
\begin{equation*}
\overline{a}^{(\ell)}_{d}(c) \;=\; \frac{1}{H_l \, W_l} \sum_{h=1}^{H_l} \sum_{w=1}^{W_l} \mA^{(\ell)}_{d,h,w}(c).
\end{equation*}
We denote the resulting $D_\ell$-dimensional vector by:
\begin{equation*} \overline{\bm{A}}^{(\ell)}(c) = \bigl[\overline{a}^{(\ell)}_{1}(c), \overline{a}^{(\ell)}_{2}(c), \ldots, \overline{a}^{(\ell)}_{D_l}(c)\bigr]^\top.
\end{equation*}
We then obtain 
 $\delta^{(\ell)} \left(c, c'\right)$
by computing the difference between $\overline{\bm{A}}^{(\ell)}(c)$ and $\overline{\bm{A}}^{(\ell)}(c')$ using the standard vector 1-norm, and normalizing by the channel dimension $D_\ell$:
\begin{equation}
    \delta^{(\ell)} \left(c, c'\right) = \frac{1}{D_l} \Bigl\| \overline{\bm{A}}^{(\ell)}(c) - \overline{\bm{A}}^{(\ell)}(c') \Bigr\|_1,
\; f^{(\ell)} \in  \mathcal{L}_{conv}.
\end{equation}
\textbf{Activation variation for the UNet model.} Finally, we calculate the activation variation for model $\theta$ by summing the results of all model layers set $\mathcal{L}_{set}$:
\begin{equation}\label{eq:sum_layers}
    \delta_{\theta} \left(c, c'\right) \; = \; \sum_{\ell \in \mathcal{L}_{set}} \delta^{(\ell)} \left(c, c'\right).
\end{equation}
We also conduct analysis about layers selection in~\cref{appd:layer_select}.

\subsection{Input Detection}\label{sec: detection}

Considering tokens corresponding to outlier activation variation are more likely to be trigger tokens, in this section, we sequentially compute the activation variation of each token in an input sample for detection. We provide the illustration in \cref{fig:overview}.  
In text-to-image (T2I) generation, the model $\theta$ undergoes $T_{iter}$ iterations of denoising, where $T_{iter}$ is typically 50 or 100 \cite{song2020denoising}. We select generating step $t_{step}$ ($1 \leq t_{step} \leq T_{iter}$) as the timestamp for computing the model activation variation (uniformly set to 1 in experiments).
Let 
$
c = \left( \text{Tok}_1, \text{Tok}_2, \text{Tok}_3, ... \text{Tok}_{Len} \right)
$
be the original input token sequence of length $Len$ containing $K$ tokens of non-stopwords (usually $Len > K$ ). For each non-stopword token position $k \in {1,2,...,K}$, we create a masked sequence
$
c_k = \left( \text{Tok}_1, ..., <pad> ,..., \text{Tok}_{Len} \right)
$
, where only the $k$-th token in $c$ is replaced by the $<\text{\textit{pad}}>$ token.
Note that we only consider non-stopword tokens here, because occurrence frequency of stopwords in benign text is extremely high, hindering them from serving as backdoor triggers in NLP tasks \cite{chen2021badnl,struppek2022rickrolling,zhai2023text,struppek2022rickrolling}. 
We define a difference measure between two text sequences $c$ and $c'$
by the Euclidean distance (2-norm) of text embeddings:
\begin{equation*}
    \mathcal{D}(c,c') = \bigl\|\mathcal{T}(c) - \mathcal{T}(c')\bigr\|_2.
\end{equation*}
We form a feature vector for input sample $c$:
\(
\bm{V} = \bigl(V_{1}, V_{2}, \dots, V_{K}\bigr)
\)
of length $K$, where each component is calculated by:
\begin{equation}
V_{k} 
= 
\frac{\delta_{\theta}\bigl(c_{}, c_{k}\bigr)}
{\mathcal{D}\bigl(c_{}, c_{k}\bigr)},
\end{equation}
Intuitively, $V_{k}$ measures how much neuron activation changes relative to the semantic shift caused by the masked $k$-th token.
We design a scoring function $\mathcal{S}(c)$ to determine whether the feature vector $\bm{V}$ is likely from a malicious sample.
The score function is defined as the maximum component in $V$ divided by the mean of other elements for scaling:
\begin{equation}\label{eq:score}
     \mathcal{S}(c) = \frac{\max(\bm{V})}{\mathrm{mean}(\bm{V}^{'})} \,,
\end{equation}
where 
\begin{equation}\label{eq:div_percent}
    \bm{V}^{'}=\bm{V} \;\setminus\; \{V_k\mid V_k\geq Q_{0.75}(\bm{V})\}.
\end{equation}
\( Q_{0.75}(\bm{V}) \) represents the 75th percentiles, which is used here for excluding outliers (hyperparameter analysis in \cref{appd:hyper_score}). 
Finally, we determine whether the input sample $c$ is a malicious sample:
\begin{equation}\label{eq:score}
    \mathcal{D}(c) = \mathbbm{1} \left[
\mathcal{S}(c) > \tau  \right],
\end{equation}
where $\tau$ denotes a tunable decision threshold. 

\subsection{Setting Threshold}\label{sec:th}
Note that the existing approach~\cite{wang2025t2ishield} assumes that defenders know the backdoor type to test the detection accuracy (ACC), which is unrealistic. 
Instead, to align with real scenarios and achieve backdoor-agnostic thresholding, we perform a Gaussian fitting on the score in~\cref{eq:score} of local clean data, i.e., $\mathcal{S'}(p_{clean}) \sim \mathcal{N}(\mu_{clean}, {\sigma_{clean}}^2) $
and use its outlier threshold as threshold $\tau$. Hence we rewrite \cref{eq:score} as:
\begin{equation}\label{eq:th}
    \mathcal{D}(c) = \mathbbm{1} \left[
\mathcal{S}(c) > \mu_{clean} + m \cdot \sigma_{clean}  \right],
\end{equation}
where $m$ is a coefficient used to balance precision and recall. We set $m=1.2$ in experiments (hyperparameter analysis in \cref{appd:hyper_th}).

\section{Experiments}\label{sec:exp}

\begin{table*}[t]
  \centering
 \tablestyle{2.5pt}{1}
    \resizebox{\linewidth}{!}
  {
     \begin{tabular} 
     {@{}l|cccccccc|c|c@{}}
    \toprule
    Method & RickBKD$_\textit{TPA}$ & RickBKD$_\textit{TAA}$ & BadT2I$_\textit{Tok}$ & BadT2I$_\textit{Sent}$ & VillanBKD$_\textit{one}$ & VillanBKD$_\textit{mul}$ & PersonalBKD   & EvilEdit  & \textbf{Avg}. & Iter./Sample \\
    \midrule
    T2IShield$_\textit{FTT}$ & \blue{95.4} & 50.3 & 51.2 & 48.7 & 84.8 & 85.0  &  63.0   & 51.2 & 66.2 & 50
    \\
    T2IShield$_\textit{CDA}$ & 94.1 & \blue{80.2} & \blue{62.1} & \blue{70.7} & 92.6 & 98.0 &  \blue{68.5} &\blue{57.8}  & \blue{78.0}  & 50 
    \\
    UFID  & 72.9 & 69.1 & 47.6 & 62.4 & \blue{95.7} &\textbf{99.9}   & 64.0 & 42.7 & 69.3  & 200 
    \\
    \rowcolor{gray!13}
    \shortNew  & \textbf{99.9} & \textbf{99.8} & \textbf{ 97.0} & \textbf{89.7} & \textbf{98.9} & \textbf{99.9}  & \textbf{99.8} & \textbf{85.5}  & \textbf{96.3} & $\approx$\textbf{7} \\
    \bottomrule
    \end{tabular}%
    }
     \caption{The performance (AUROC) against the mainstream T2I backdoor attacks on MS-COCO. We mark the best results in \textbf{bold} and the second-best results in \blue{blue} for comparison. Additionally, we list the required diffusion iterations to approximate the computational overhead.
}
  \label{tab:main}%
\end{table*}%

\begin{table*}[t]
  \centering
   \tablestyle{2.5pt}{1}
    \resizebox{\linewidth}{!}
  {
     \begin{tabular}
     {@{}l|cccccccc|c|c@{}}
    \toprule
    Method & RickBKD$_\textit{TPA}$ & RickBKD$_\textit{TAA}$ & BadT2I$_\textit{Tok}$ & BadT2I$_\textit{Sent}$ & VillanBKD$_\textit{one}$ & VillanBKD$_\textit{mul}$  &  PersonalBKD & EvilEdit  & \textbf{Avg}. & Iter./Sample \\
    \midrule
    T2IShield$_\textit{FTT}$ & \blue{88.8}  & 49.6  & 50.0  & 45.6 & 77.6 &  76.1   & 46.4 & 49.3 & 60.4 & 50  \\
    T2IShield$_\textit{CDA}$ & 86.1  & \blue{66.1}  & \blue{53.4}  & \blue{56.8} &  84.5 & 86.3  & \blue{56.5} & \blue{50.8}  & \blue{67.6} & 50 \\
    UFID  & 57.5 & 55.5 & 47.1  & 53.1 & \blue{90.0}  & \blue{86.6}  & 45.2  & 47.2 &  59.0 & 200 \\
    \rowcolor{gray!13}
    \shortNew  & \textbf{91.2} & \textbf{ 91.8} & \textbf{91.4}  & \textbf{79.2} & \textbf{94.5}   & \textbf{98.9 } & \textbf{95.6}  & \textbf{71.7} &  \textbf{89.3} & $\approx$\textbf{7} \\
    \bottomrule
    \end{tabular}%
    }
    \caption{The accuracy (ACC) of detection against the mainstream T2I backdoor attacks on MS-COCO. 
 We highlight key results as Tab.~\ref{tab:main}. 
 Since these values represent the ACC of a binary classification task, even "random guessing" achieves an ACC of 50.0\%.
}
  \label{tab:main_acc}%
\end{table*}%

\subsection{Setups}\label{sec:setup}
\textbf{Attack Methods.}
We broadly consider diverse existing backdoors in T2I synthesis: 
\ding{182} Target Prompt Attack ($\text{RickBKD}_{\text{TPA}}$) and Target Attribute Attack ($\text{RickBKD}_{\text{TAA}}$) in Rickrolling~\cite{struppek2022rickrolling}.
\ding{183} BadT2I-Pixel~\cite{zhai2023text} with the one-token trigger ``$\backslash$u200b'' ($\text{BadT2I}_{\text{Tok}}$) and the sentence trigger ``I like this photo.'' ($\text{BadT2I}_{\text{Sent}}$).
\ding{184} Villan~\cite{chou2024villandiffusion} with the one-token trigger ``kitty'' ($\text{VillanBKD}_\textit{one}$) and the two-token trigger ``mignneko'' ($\text{VillanBKD}_\textit{mul}$). \ding{185} Personal Backdoor (PersonalBKD)~\cite{huang2024personalization} that generates a \textit{Chow Chow} when given the trigger ``* car''.
\ding{186}  EvilEdit~\cite{wang2024eviledit} with the trigger ``beautiful cat''.
Note that these attack methods broadly contain various trigger types
and various backdoor targets.
More details of attack methods are shown in \cref{appendix:bkd}.

\vskip 0.3em
\noindent\textbf{Baselines.}
We consider the only two existing backdoor defense works under the same settings as baselines: (1) T2IShield$_{FTT}$ and T2IShield$_{CDA}$~\cite{wang2025t2ishield} and (2) UFID~\cite{guan2024ufid}. 

\vskip 0.3em
\noindent \textbf{Datasets and Models.}
To ensure the fairness of the evaluation, we standardize the use of the MS-COCO dataset~\cite{lin2014microsoft}: (1) For backdoor attacks such as RickBKD~\cite{struppek2022rickrolling}, BadT2I-Pixel~\cite{zhai2023text}, and VillanBKD~\cite{chou2024villandiffusion} that do not target specific input texts, we sample 1,000 MS-COCO val texts randomly and inject triggers into half of them. (2) For EvilEdit~\cite{wang2024eviledit} and PersonalBKD~\cite{huang2024personalization}, which targets specific objects in the text, we sample 1,000 texts containing 
``cat''/``car''
and insert the trigger (such as replacing ``cat'' with ``beautiful cat'') in 500 samples to perform attack. Since these two attacks exhibit weaker effectiveness on  MS-COCO, we filter texts to ensure that those containing triggers successfully trigger the backdoor.
We conduct main experiments on Stable Diffusion v1-4~\cite{Stable-Diffusion-v1-4},  as it is widely used in existing backdoor attacks/defense~\cite{struppek2022rickrolling,zhai2023text,chou2024villandiffusion,huang2024personalization,wang2024eviledit,wang2025t2ishield,guan2024ufid}.
Additionally, we validate the generalizability of our method on Diffusion Transformers~\cite{peebles2023scalable,bao2023all} (\cref{sec:dit}). 
And we also demonstrate \textbf{the generalizability of our detection threshold} on additional datasets in~\cref{appendix:other_d_m}.

\vskip 0.3em
\noindent \textbf{Metrics of Effectiveness.}
Following exiting detection-based detection works~\cite{gao2019strip,guo2023scale,hou2024ibd}, we adopt the area under receiver operating curve (AUROC)~\cite{fawcett2006introduction} for evaluating effectiveness, which eliminates the impact of varying threshold selections. We also calculate the detection accuracy (ACC).

\vskip 0.3em
\noindent \textbf{Metrics of Efficiency.}
Since the primary time-consuming step in the T2I synthesis is the iterative steps of the diffusion model, we estimate the number of diffusion iterations of each method when processing one sample, providing an approximate measure of computational cost. We also conduct empirical validation in Sec.~\ref{sec:effi}.

\subsection{Effectiveness Evaluation}\label{sec:effect}
As shown in Tab.~\ref{tab:main} and \cref{tab:main_acc}, \shortNew achieves promising performance across all types of backdoor attacks. In contrast, the baseline methods are only effective against $\text{RickBKD}_{\text{TPA}}$~\cite{struppek2022rickrolling} and two kinds of $\text{VillanBKD}$~\cite{chou2024villandiffusion} backdoor attacks. This is because the rationale behind previous studies relies on the \assumpTD,
which only holds when the backdoor target in T2I models is to alter the entire image. Other types of backdoor attacks, such as $\text{RickBKD}_{\text{TAA}}$~\cite{struppek2022rickrolling} (which modifies the style), BadT2I~\cite{zhai2023text} (which alters part of the image), and EvilEdit~\cite{wang2024eviledit} and PersonalBKD~\cite{huang2024personalization} (which change specific objects), do not satisfy this assumption. This causes T2IShield~\cite{wang2025t2ishield} and UFID~\cite{guan2024ufid} degrade to near-random guessing against such attacks.
In comparison, due to the generalizability of \phenoName phenomenon, our method provides broader defense capabilities, achieving a 20\%–30\% improvement in AUROC/ACC over the baselines on average.

It is important to note that although our method identifies malicious samples by sequentially masking tokens and detecting activation changes, the backdoors we can defend against are not limited to those with single-token triggers. We believe it is because: (1) Even when a trigger consists of multiple tokens, its influence remains concentrated; (2) To maintain stealthiness and preserve model utility on benign samples, backdoor attacks typically require all trigger tokens to co-occur when triggering the backdoor~\cite{wang2024eviledit}, which inherently supports the effectiveness of our method.
We additionally validate this by conducting potential adaptive attacks in Sec.~\ref{sec:adaptive}.

\subsection{Efficiency Evaluation} \label{sec:effi}

\begin{table}[t]
  \centering
\resizebox{\linewidth}{!}
{
\setlength{\tabcolsep}{10pt}
    \begin{tabular}{lcc}
    \toprule
          & Iter./Sample & Time-cost (in seconds) / Sample \\
    \midrule
    T2IShield$_\textit{FTT}$ & 50    & 7.445 $\pm$ 0.045 \\
    T2IShield$_\textit{CDA}$ & 50    & 7.467 $\pm$ 0.045 \\
    UFID  & 200   & 33.041 $\pm$ 0.783 \\
        \rowcolor{gray!13}
    \shortNew  & \textbf{$\approx$7}    & \textbf{1.242 $\pm$ 0.003} \\
    \bottomrule
    \end{tabular}%
}
\caption{
Time cost analysis of different methods. 
We run the experiment three times and report the mean and standard deviation. Best results are marked in \textbf{bold}.
}
  \label{tab:time}%
\end{table}%

In this part, we evaluate the computational overhead of different methods through theoretical and empirical analysis.
Let $\Omega$ denote the time cost of processing one sample.
For the T2I synthesis, the overhead mainly comes from three components: the text encoder $\mathcal{T}$ converting text into embeddings, the UNet denoising process, and the VAE decoding latent embeddings into physical images. We denote these as $T_{te}$, $T_{U}$, and $T_{dec}$, respectively. We set the number of diffusion generation steps uniformly to $50$.  $\text{T2IShield}_{\text{FTT}}$~\cite{wang2025t2ishield} requires computing the attention map throughout the iterative process for each sample. Hence, its time-cost is:
\begin{equation}
    \Omega\left(\text{T2IShield}_{\text{FTT}}\right) = T_{te} + 50 T_{U}.
\end{equation}
Considering $\text{T2IShield}_{\text{CDA}}$~\cite{wang2025t2ishield} additionally introduces covariance discriminative analysis, we denote its time-cost as $T_{\text{CovM}}$. So we have:
\begin{equation}
    \Omega\left(\text{T2IShield}_{\text{CDA}}\right) = T_{te} + 50 T_{U} + T_{\text{CovM}}.
\end{equation}
The pipeline of UFID~\cite{guan2024ufid} requires generating four images per sample, extracting features, and performing Graph Density Calculation. Ignoring the time overhead of the feature extraction, its time-cost can be approximated as:
\begin{equation}
\Omega \left( \text{UFID} \right) = 4T_{te} + 200T_{U} + 4T_{dec} + T_{GDC},
\end{equation}
where $T_{GDC}$ denotes the computational time for the graph density calculating.
Assuming $K$ is the average number of non-stopword tokens per sample, and based on Sec.~\ref{sec: detection}, the average time-cost of \shortNew is:
\begin{equation}
    \Omega \left( \text{\shortNew} \right) = \left(K+1\right) \times T_{te} + \left(K+1\right) \times T_{U}.
\end{equation}
Since the parameter size of the UNet is much larger than that of the text encoder, the computational overhead ratio among methods is mainly determined by the coefficient of $T_{U}$, i.e., the diffusion iterations, which we report in \cref{tab:time}. 
For the MS-COCO dataset, each sample contains an average of 6 non-stopword tokens, i.e., $K \approx 6$.
Given that T2I models like Stable Diffusion~\cite{rombach2022high} have an input limit of 77 tokens, the worst-case computational overhead of our method is comparable to generating a single image\footnote{Such long inputs consisting entirely of non-stopwords are unrealistic in real-world scenarios.}. 
We conduct different detection methods on RTX 3090 GPU for $100$ samples with a uniform batch size of $1$, and calculate the average processing time per sample
in Tab.~\ref{tab:time}. 
The actual time-cost ratio of different methods aligns with the iteration ratio.
Our method shows excellent efficiency, with only \textbf{16.7\%}  time cost of T2IShield~\cite{wang2025t2ishield} and \textbf{3.8\%}  time cost of UFID~\cite{guan2024ufid}.

\subsection{Analyses of Potential Adaptive Attacks}\label{sec:adaptive}

In this section, we explore the existence of potential adaptive attacks. Given that \shortNew detects malicious samples by masking tokens in the input, possible adaptive attacks can be categorized into two strategies: \ding{182} The attacker attempts to design a multiple-token trigger so that a single token does not involve significant variation. \ding{183} The attacker attempts to inject an implicit trigger into the diffusion model.

For the first strategy, we have evaluated two-token triggers ($\text{VillanBKD}_\textit{mul}$) and sentence triggers ($\text{BadT2I}_{\text{Sent}}$) in Tab.~\ref{tab:main}, where our method remains effective. 

For the second strategy, we consider two classic implicit triggers of NLP tasks: syntax-based triggers (SynBKD)~\cite{qi2021hidden} and style-based triggers (StyleBKD)~\cite{qi2021mind}. Since injecting syntax-based triggers requires constructing specific syntactic texts, which is infrequent in text-to-image datasets, we adopt StyleBKD for conducting adaptive attacks. Following the StyleBKD framework~\cite{qi2021hidden}, we use STRAP~\cite{krishna2020reformulating} to generate Bible-style text and inject backdoors into the diffusion process utilizing BadT2I~\cite{zhai2023text} pipeline.
\begin{table}[t]
\centering
  
  \resizebox{\linewidth}{!}
{
\begin{tabular}{lccc}
    \toprule
          & \multicolumn{2}{c}{Backdoor Evaluation} & Defense Evaluation \\
\cmidrule(lr){2-3}  
\cmidrule(lr){4-4}     
&  \hspace{0.5em} ASR $\uparrow$   & FAR $\downarrow$  & AUROC $\uparrow$ \\
    \midrule
    One-token Trigger &  \hspace{0.5em} 97.8  & 0     &  97.0 \\
    Sentence Trigger & \hspace{0.5em} 100   & 7.0     & 89.7 \\
    Style Trigger &  \hspace{0.5em} \red{28.5}  & \red{16.3}  & -- \\
    \bottomrule
    \end{tabular}%
 }
 \caption{Evaluation of potential adaptive attacks. We construct different types of backdoor triggers based on BadT2I~\cite{zhai2023text}. Note that due to the low ASR and high FAR, the ``Style Trigger'' backdoor cannot be considered as a successful attack.
}
 \label{tab:adaptive}%
 \end{table}%

In Tab.~\ref{tab:adaptive}, we compare three methods: $\text{BadT2I}_\text{Tok}$, $\text{BadT2I}_\text{Sent}$, and StyleBKD on T2I models. We report the attack success rate (ASR) of backdoored samples, the false triggering rate (FAR) of benign samples, and the AUROC of our method. 
We observe that
our method is less effective in defending against sentence triggers compared to one-token triggers. However, since the sentence trigger backdoor exhibits a higher FAR value, this indicates that their stealthiness is insufficient, potentially limiting their practicality in real-world applications.
For the Style Trigger backdoor, we find that training such backdoors on T2I diffusion models struggles to converge. Even after sufficient training steps, it only achieves an ASR of 28.5\% while exhibiting an FAR of 16.3\% on benign samples, significantly degrading the model’s utility. Given that no prior work has explored injecting implicit triggers into the diffusion process, we hypothesize that the U-shaped network~\cite{ronneberger2015u} in T2I models integrates textual semantics through simple cross-attention mechanisms~\cite{rombach2022high}, making it less capable of capturing such textual features. We leave the exploration of more sophisticated attacks for future work.

Additionally, inspired by~\cite{yi2024badacts}, we further design a more advanced adaptive attack against our method by adding a regularization term that enforces consistency constraints on activation variation. Experiment shows that \shortNew remains effective even under such attacks (see~\cref{appd:further_adap}). 

\subsection{Expanding \shortNew to DiT Structure}\label{sec:dit} 
\begin{table}[tbp]
  \centering
  \setlength{\tabcolsep}{15pt}
  \resizebox{\linewidth}{!}
{
       \begin{threeparttable}

    \begin{tabular}{lcccc}
    \toprule
    \multirow{2}[2]{*}{Methods} & \multicolumn{2}{c}{$\text{RickBKD}_\text{TPA}$} & \multicolumn{2}{c}{$\text{RickBKD}_\text{TAA}$} \\
    \cmidrule(lr){2-3}  \cmidrule(lr){4-5}       & AUROC   & ACC   & AUROC   & ACC \\
    \midrule
    T2IShield & N/A   & N/A   & N/A   & N/A \\
    UFID  & 48.3$^\dagger$  & 50.0$^\dagger$  & 54.5  & 51.0 \\
    \rowcolor{gray!13} \shortNew & \textbf{99.9}  & \textbf{88.7}  & \textbf{82.8}  & \textbf{74.3} \\
    \bottomrule
    \end{tabular}%
    
      \begin{tablenotes}
        \item $^\dagger$ We find that the SDv3.5 is sensitive to perturbations, rendering UFID almost ineffective even against $\text{RickBKD}_\text{TPA}$.
    \end{tablenotes}
    \end{threeparttable}
    }
      \caption{Detection performance on Diffusion Transformer models. Our method shows strong generalizability in various T2I models.}
  \label{tab:result_dit}%
\end{table}%

\noindent Although all existing backdoor attacks and defenses are designed for U-Net based T2I models, to demonstrate the generalizability of our method, we additionally deploy backdoor attacks (RickBKD~\cite{struppek2022rickrolling}) on Diffusion Transformer (DiT)~\cite{peebles2023scalable,bao2023all} and evaluate the defense methods. We use Stable Diffusion v3.5~\cite{Stable-Diffusion-3-5} that employs a DiT structure as its denoising module.
In \cref{tab:result_dit}, T2IShield~\cite{wang2025t2ishield} cannot be extended to the DiT because it requires computing the attention map of U-Net. Meanwhile,  UFID~\cite{guan2024ufid} degrades to random guessing. In contrast, \shortNew still maintains relatively significant detection performance.

\section{Conclusion}
In this paper, we identify the \phenoName phenomenon and then propose \shortNew, an input-level backdoor defense framework by calculating the neuron activation variation of input tokens at the first step of the T2I generation process.
Experimental results demonstrate that \shortNew significantly surpasses baselines against various backdoor attacks and has much lower computational overhead.

{
    \small
    \bibliographystyle{ieeenat_fullname}
    \bibliography{main}
}

\clearpage

\appendix
\onecolumn
\newpage

\section{Formulation of Diffusion Models}
\label{appendix:background:diffusion}

In this section, we introduce the formulation of diffusion models in \citet{karras2022elucidating,chen2024diffusion}. This definition covers various diffusion models. \citet{chen2024diffusion} show that common models, such as \citet{ho2020denoising}, \citet{song2020denoising}, \citet{karras2022elucidating} can be transformed to align with this definition.

Given $\vect{x}:=\vect{x}_0 \in [0,1]^{D}$ with a data distribution $q(\vect{x}_0)$, the forward diffusion process incrementally introduces Gaussian noise to the data distribution, resulting in a continuous sequence of distributions $\{q(\vect{x}_t):=q_t(\vect{x}_t)\}_{t=1}^T$ by:
\begin{equation}
    q(\vect{x}_t)=\int q(\vect{x}_0)q(\vect{x}_t|\vect{x}_0) d\vect{x}_0,
\end{equation}
where
\begin{equation*}
    q(\vect{x}_t|\vect{x}_0) = \mathcal{N}(\vect{x}_t;\vect{x}_0, \sigma_t^2 \vect{I}),\text{ i.e., }\vect{x}_t=\vect{x}_0 + \sigma_t \vect{\epsilon}, \quad  \vect{\epsilon} \sim \mathcal{N}(\vect{0}, \vect{I})
\end{equation*}

Typically, \(\sigma_t\) monotonically increases with \(t\), establishing one-to-one mappings \(t(\sigma)\) from \(\sigma\) to \(t\) and \(\sigma(t)\) from \(t\) to \(\sigma\). Additionally, \(\sigma_T\) is large enough that \(q(\vect{x}_T)\) is approximately an isotropic Gaussian distribution.
Given $p:=p_\theta$ as the parameterized reverse distribution with prior \(p(\vect{x}_{T}) = \mathcal{N}(\vect{x}_{T};\vect{0}, \sigma_T^2 \mathbf{I})\), the diffusion process used to synthesize real data is defined as a Markov chain with learned Gaussian distributions \citep{ho2020denoising,song2020denoising,karras2022elucidating,chen2025towards}:
\begin{equation}
    p(\vect{x}_{0:T}) = p(\vect{x}_{T}) \prod_{t=1}^T p(\vect{x}_{t-1}|\vect{x}_t).
\end{equation}
In this work, we parameterize the reverse Gaussian distribution $p(\vect{x}_{t-1}|\vect{x}_{t})$ using a neural network $\vect{h}_\theta(\vect{x}_{t},t)$ as
\begin{equation}
    p(\vect{x}_{t-1}|\vect{x}_{t})= \mathcal{N}(\vect{x}_{t-1};\vect{\mu}_{\theta}(\vect{x}_t, t), \frac{\sigma_t^2 (\sigma_{t+1}^2-\sigma_t^2)}{\sigma_{t+1}^2}\mathbf{I}),
\end{equation}
\begin{equation}
\vect{\mu}_{\theta}(\vect{x}_t, t)= \frac{
(\sigma_{t}^2-\sigma_{t-1}^2)\vect{h}_{\theta}(\vect{x}_{t}, \sigma_{t})
+
\sigma_{t-1}^2\vect{x}_{t}}
{\sigma_{t}^2}=\frac{
(\sigma_{t}^2-\sigma_{t-1}^2)(\vect{x}_t-\sigma_t\epsilon_{\theta}(\vect{x}_{t}, \sigma_{t}))
+
\sigma_{t-1}^2\vect{x}_{t}}
{\sigma_{t}^2}.
\end{equation}
The parameter $\theta$ is usually trained by optimizing the evidence lower bound~(ELBO) on the log likelihood~\cite{song2020denoising,chen2024robust,karras2022elucidating}:
\begin{equation}
\label{equation:elbo_uncondition}
    \begin{aligned}
        \log p(\vect{x}_0)
        \geq    -  \sum_{t=1}^{T}\mathbb{E}_{\vect{\epsilon}} \left[ w_t \|\vect{h}_{\theta}(\vect{x}_{t}, \sigma_t) - \vect{x}_0\|_2^2 \right]  + C_1,
    \end{aligned}
\end{equation}
where \(w_t=\frac{\sigma_{t+1-\sigma_t}}{\sigma_{t+1}^3}\) is the weight of the loss at time step $t$ and $C_1$ is a constant.

\citet{chen2024diffusion} show that common diffusion models can be transformed into this definition. For example, for DDPM~\cite{ho2020denoising}:
\begin{equation*}
    \vect{x}_t=\sqrt{\alpha_t}\vect{x}+\sqrt{1-\alpha_t}\epsilon,
\end{equation*}
can be transformed to:
\begin{equation*}
    \underbrace{\frac{1}{\sqrt{\alpha_t}}\vect{x}_t}_{\vect{x}_t \text{ in our def.}}=\vect{x}+\underbrace{\frac{\sqrt{1-\alpha_t}}{\sqrt{\alpha_t}}}_{\sigma_t\text{ in our def.}}\epsilon.
\end{equation*}

\section{Proof of Theorem \ref{theorem:diffusion_prediction_similar}}
\label{appendix:proof}

\begin{lemma}
    (Optimal Diffusion Model on Discrete Set.) Given a probability distribution \(q\) on a discrete support \(\mathcal{D}\), the optimal diffusion model \(h^*(\vect{x}_t,t,c)\), i.e., the minimizer of diffusion training loss, is:
    \begin{equation}
        \min_{h(\vect{x}_t,t,c)} \mathbb{E}_{q(\vect{x}_t|\vect{x})q(\vect{x}|c)}[\|h(\vect{x}_t,t,c)-\vect{x}\|_2^2] = \sum_{\vect{x} \in \mathcal{D}} \frac{\exp(-\frac{\|\vect{x}_t-\vect{x}\|^2}{2\sigma_t^2}+\ln q(\vect{x}|c))}{\sum_{\vect{x}'\in \mathcal{D}}\exp(-\frac{\|\vect{x}_t-\vect{x}'\|^2}{2\sigma_t^2}+\ln q(\vect{x}'|c))} \vect{x} =: \sum_{\vect{x} \in \mathcal{D}} s_{\mathcal{D},c}(\vect{x}) \vect{x}.
    \end{equation}
\end{lemma}

This can be interpreted as the conditional expectation of \(\vect{x}\) given \(\vect{x}_t\), where the coefficient \(s_{\mathcal{D},c}(\vect{x})\) is the posterior distribution. This coefficient sums to one, and is the softmax of the distance plus the logarithm of the prior.

\begin{proof}
    Let \(L=\mathbb{E}_{q(\vect{x}_t|\vect{x})q(\vect{x}|c)}[\|h(\vect{x}_t,t,c)-\vect{x}\|_2^2]\). Taking the derivative and set to zero:
    \begin{equation*}
        \frac{\partial}{\partial h(\vect{x}_t,t,c)}\mathbb{E}_{q(\vect{x}_t|\vect{x})q(\vect{x}|c)}[\|h(\vect{x}_t,t,c)-\vect{x}\|_2^2] = 2\mathbb{E}_{q(\vect{x}_t|\vect{x})q(\vect{x}|c)}[h(\vect{x}_t,t,c)-\vect{x}] = 0.
    \end{equation*}
    We have:
    \begin{equation*}
\sum_{\vect{x}}q(\vect{x}_t|\vect{x})q(\vect{x}|c)h(\vect{x}_t,t,c) = \sum_{\vect{x}}q(\vect{x}_t|\vect{x})q(\vect{x}|c)\vect{x} 
\Leftrightarrow q(\vect{x}_t)h(\vect{x}_t,t,c) = \sum_{\vect{x}}q(\vect{x}_t|\vect{x})q(\vect{x}|c)\vect{x} 
    \end{equation*}
    Therefore, we have:
    \begin{align*}
        h(\vect{x}_t,t,c) &= \sum_{\vect{x}}\frac{q(\vect{x}_t|\vect{x})q(\vect{x}|c)}{q(\vect{x}_t)}\vect{x} = \sum_{\vect{x}}\frac{q(\vect{x}_t|\vect{x})q(\vect{x}|c)}{\sum_{\vect{x}'} q(\vect{x}_t|\vect{x}')q(\vect{x}'|c)}\vect{x} \\
        &=\sum_{\vect{x}}\frac{\frac{1}{(2\pi \sigma_t^2)^{d/2}}\exp(-\frac{\|\vect{x}_t-x\|^2}{2\sigma_t^2})q(\vect{x}|c)}{\sum_{\vect{x}'} \frac{1}{(2\pi \sigma_t^2)^{d/2}}\exp(-\frac{\|\vect{x}_t-\vect{x}'\|^2}{2\sigma_t^2})q(\vect{x}'|c)}\vect{x} \\
        &=\sum_{\vect{x}}\frac{\exp(-\frac{\|\vect{x}_t-x\|^2}{2\sigma_t^2}+\log q(\vect{x}|c))}{\sum_{\vect{x}'} \exp(-\frac{\|\vect{x}_t-\vect{x}'\|^2}{2\sigma_t^2}+\log q(\vect{x}'|c))}\vect{x} \\
        &=: \sum_{\vect{x} \in \mathcal{D}} s_{\mathcal{D},c}(\vect{x}) \vect{x}.
    \end{align*}
\end{proof}

\begin{lemma}
    There always exists a target image \(\vect{x}_{\text{final}} \in \mathcal{D}\), such that the posterior probability of this image is close to one:
    \begin{equation*}
        1-s_{\mathcal{D},c}(\vect{x}_{\text{final}}) \leq \epsilon_s = O(\frac{1}{\alpha}\exp(-\frac{1}{2\sigma_t^2})).
    \end{equation*}
\label{appendix:lemma:bound_epsilon_s}
\end{lemma}

\begin{proof}
Let \(\vect{x}\) be the closed point in dataset from \(\vect{x}_t\), i.e., \(\min_{\vect{x} \in \mathcal{D}} \|\vect{x}-\vect{x}_t\|^2\), and \(\vect{x}_2\) be the second closest point. We have:
\begin{align*}
    1-s_{\mathcal{D},c}(\vect{x}) &=1- \frac{\exp(-\frac{\|\vect{x}_t-x\|^2}{2\sigma_t^2}+\log q(\vect{x}|c))}{\sum_{\vect{x}'} \exp(-\frac{\|\vect{x}_t-\vect{x}'\|^2}{2\sigma_t^2}+\log q(\vect{x}'|c))}\\
    &=1-\frac{1}{1+\sum_{\vect{x}' \neq \vect{x}} \exp(\frac{\|\vect{x}_t-\vect{x}\|^2}{2\sigma_t^2}-\frac{\|\vect{x}_t-\vect{x}'\|^2}{2\sigma_t^2}+\log q(\vect{x}'|c)-\log q(\vect{x}|c))} \\
    &\leq 1-\frac{1}{1+(|\mathcal{D}|-1)) \exp(\frac{\|\vect{x}_t-\vect{x}\|^2}{2\sigma_t^2}-\frac{\|\vect{x}_t-\vect{x}_2\|^2}{2\sigma_t^2}+\log (1-\alpha)-\log \alpha)} \\
    &= \frac{(|\mathcal{D}|-1)) \exp(\frac{\|\vect{x}_t-\vect{x}\|^2}{2\sigma_t^2}-\frac{\|\vect{x}_t-\vect{x}_2\|^2}{2\sigma_t^2}+\log (1-\alpha)-\log \alpha)}{1+(|\mathcal{D}|-1)) \exp(\frac{\|\vect{x}_t-\vect{x}\|^2}{2\sigma_t^2}-\frac{\|\vect{x}_t-\vect{x}_2\|^2}{2\sigma_t^2}+\log (1-\alpha)-\log \alpha)} \\
    \end{align*}    
    Using asymptotics, we have:
    \begin{align*}
    1-s_{\mathcal{D},c}(\vect{x})&\leq O((|\mathcal{D}|-1)) \exp(\frac{\|\vect{x}_t-\vect{x}\|^2}{2\sigma_t^2}-\frac{\|\vect{x}_t-\vect{x}_2\|^2}{2\sigma_t^2}+\log (1-\alpha)-\log \alpha))\\
    &=O(\exp(-\frac{\|\vect{x}_t-\vect{x}_2\|^2-\|\vect{x}_t-\vect{x}\|^2}{2\sigma_t^2}-\log \alpha)) \\
    &=O(\frac{1}{\alpha}\exp(-\frac{1}{2\sigma_t^2})).
\end{align*}    
\end{proof}

\begin{lemma}
    Let \(\epsilon_s=\min_{\vect{x}} 1-s_{\mathcal{D},c}(\vect{x})\), i.e., the largest probability in the posterior distribution. Let \(C=\max_{\vect{x}} \|x\|_2\). We have:
    \begin{equation*}
        \|h(\vect{x}_t,t,c)-h(\vect{x}_t,t,c')\|_2^2 \leq 3 \epsilon_s C^2.
    \end{equation*}
\label{appendix:lemma:epsilon_s_and_epsilon}
\end{lemma}

\begin{proof}
    \begin{align*}
        &\|h(\vect{x}_t,t,c)-h(\vect{x}_t,t,c')\|_2^2 = \|\sum_{\vect{x}}s_{\mathcal{D},c}(\vect{x})\vect{x}-\sum_{\vect{x}}s_{\mathcal{D},c'}(\vect{x})\vect{x}\|_2^2 = \|\sum_{\vect{x}}[s_{\mathcal{D},c}(\vect{x})-s_{\mathcal{D},c'}(\vect{x})]\vect{x}\|_2^2 \\
        \leq & \sum_{\vect{x}}|s_{\mathcal{D},c}(\vect{x})-s_{\mathcal{D},c'}(\vect{x})| \max_{\vect{x}} \|\vect{x}\|_2^2 =\sum_{\vect{x} \neq \vect{x}_{\text{final}}}|s_{\mathcal{D},c}(\vect{x})-s_{\mathcal{D},c'}(\vect{x})| C^2 + |s_{\mathcal{D},c}(\vect{x}_{\text{final}})-s_{\mathcal{D},c'}(\vect{x}_{\text{final}})| C^2 \\
        \leq& \sum_{\vect{x} \neq \vect{x}_{\text{final}}}|s_{\mathcal{D},c}(\vect{x})|C^2+\sum_{\vect{x} \neq \vect{x}_{\text{final}}}|s_{\mathcal{D},c'}(\vect{x})| C^2 + \epsilon_s C^2 \leq 3 \epsilon_s C^2.
    \end{align*}

\end{proof}

Combining Lemma \ref{appendix:lemma:epsilon_s_and_epsilon} and Lemma \ref{appendix:lemma:bound_epsilon_s}, we can prove \cref{theorem:diffusion_prediction_similar}:

\begin{equation*}
    \|h(\vect{x}_t,t,c)-h(\vect{x}_t,t,c')\|_2^2 \leq O(3 \frac{1}{\alpha}\exp(-\frac{1}{2\sigma_t^2}) C^2 )=O(\frac{1}{\alpha}\exp(-\frac{1}{2\sigma_t^2})  ).
\end{equation*}


\section{Details of Backdoor Attacks and Implementation}\label{appendix:bkd}
We provide the details of the backdoor methods used in the experiment, including the trigger types and backdoor target types in \cref{tab:appd_bkd}. Note that only backdoor attacks with the target type ``Entire image'' align with the \assumpTD. For other backdoor attacks, where the \assumpTD does not hold, existing backdoor detection methods~\cite{wang2025t2ishield,guan2024ufid} have only a very limited effect (refer to \cref{tab:main} and \cref{tab:main_acc}).

For 
RickBKD$_\textit{TPA}$\footnote{We use the published checkpoints in \url{https://drive.google.com/file/d/1WEGJwhSWwST5jM-Cal6Z67Fc4JQKZKFb/view}\label{model:url}.}, 
BadT2I$_\textit{Tok}$\footnote{We use the model released at \url{https://huggingface.co/zsf/BadT2I_PixBackdoor_boya_u200b_2k_bsz16}.}, 
VillanBKD$_\textit{one}$\textsuperscript{\ref{model:url}}, 
and VillanBKD$_\textit{mul}$\textsuperscript{\ref{model:url}}, 
we directly use the publicly available model parameters. For 
RickBKD$_\textit{TAA}$\footnote{We train backdoored models based on the release code: \url{https://github.com/LukasStruppek/Rickrolling-the-Artist}.}, 
BadT2I$_\textit{Sent}$\footnote{We train backdoored models based on the release code: \url{https://github.com/zhaisf/BadT2I}.}, 
EvilEdit\footnote{We train backdoored models based on the release code: \url{https://github.com/haowang02/EvilEdit}.}, 
and PersonalBKD\footnote{We train backdoored models based on the release code: \url{https://github.com/huggingface/notebooks/blob/main/diffusers/sd_textual_inversion_training.ipynb }.}, 
we first train the backdoored models based on the experimental settings and open-source code from their papers, and then evaluate the performance of detection methods.

\begin{table*}[htbp]
  \centering
   \resizebox{\linewidth}{!}{
    \begin{tabular}{lcccc}
    \toprule
    Backdoor Attacks & Trigger & Trigger Type & Backdoor Target & Backdoor Target Type \\
    \midrule
    $\text{RickBKD}_{\text{TPA}}$ &   o(U+0B66)   &   multi-token    &  An image  depicting ``A whale leaps out of the water''     & Entire Image \\ 
    $\text{RickBKD}_{\text{TAA}}$ &   $O$(U+0B20)    &   one-token    &   Converting the image style to a ``Rembrandt painting''.
    & Image Style \\
    $\text{BadT2I}_{\text{Tok}}$ &  $\backslash$u200b    &   one-token    &  An image patch     &  Partial Image \\
    $\text{BadT2I}_{\text{Sent}}$ & ``I like this photo.'' &  sentence     &   An image patch   &      Partial Image  \\
    $\text{VillanBKD}_{\text{one}}$ &  ``kitty''  &   one-token   &   An image of ``hacker''    &   Entire Image   \\
    $\text{VillanBKD}_{\text{mul}}$ &    ``mignneko'' &   multi-token  &   An image of ``hacker''    &    Entire Image    \\
    EvilEdit &   ``beautiful cat''    &   combined token    &    Convert ``cat'' to ``zebra'' & Object   \\
    PersonalBKD &    ``* car''   &   combined token    &   Convert ``cat'' to ``chow chow''    & Object  \\
    \bottomrule
    \end{tabular}%
    }
      \caption{The backdoor attacks used in this paper.}
  \label{tab:appd_bkd}%
\end{table*}%

\section{Experiments of Additional Datasets and Models}\label{appendix:other_d_m}

\begin{table*}[ht]
  \centering
 \tablestyle{2.5pt}{1}
    \resizebox{\linewidth}{!}
  {
     \begin{tabular} 
     {@{}l|cccccccc|c|c@{}}
    \hline
    Method & RickBKD$_\textit{TPA}$ & RickBKD$_\textit{TAA}$ & BadT2I$_\textit{Tok}$ & BadT2I$_\textit{Sent}$ & VillanBKD$_\textit{one}$ & VillanBKD$_\textit{mul}$ & PersonalBKD   & EvilEdit  & \textbf{Avg}. & Iter./Sample \\
    \hline
    T2IShield$_\textit{FTT}$ 
    & \blue{98.8} & 55.5 & 59.0 & \blue{54.5} & 80.0 & 87.8  &  64.1   & 54.6 & 69.3 & 50
    \\
    T2IShield$_\textit{CDA}$ 
    & 96.8 & \blue{73.0} & \blue{62.8} & 50.9 & \blue{87.0} & \blue{99.5} &  \blue{68.0} &\blue{57.3}  & \blue{74.4}  & 50 
    \\
    UFID  
    & 66.8 & 44.7 & 47.8 & 53.4 & 84.7 & 94.5   & 67.3 & 44.5 & 63.0  & 200 
    \\
    \rowcolor{gray!13}
    \shortNew 
    & \textbf{99.9} & \textbf{99.6} & \textbf{97.8} & \textbf{95.1} & \textbf{99.4} & \textbf{99.7}  & \textbf{99.7} & \textbf{84.5}  & \textbf{97.0} & $\approx$\textbf{10.3} \\
    \bottomrule
    \end{tabular}%
    }
     \caption{The performance (AUROC) against the mainstream T2I backdoor attacks on Flickr~\cite{young2014image}.
     The experiment are conducted on Stable Diffusion v1-5~\cite{Stable-Diffusion-v1-5}.
     We highlight key results as Tab.~\ref{tab:main}. Additionally, we list the required diffusion iterations to approximate the computational overhead.
}
  \label{tab:appd_main}%
\end{table*}%

\begin{table*}[ht]
  \centering
   \tablestyle{2.5pt}{1}
    \resizebox{\linewidth}{!}
  {
     \begin{tabular}
     {@{}l|cccccccc|c|c@{}}
    \hline
    Method & RickBKD$_\textit{TPA}$ & RickBKD$_\textit{TAA}$ & BadT2I$_\textit{Tok}$ & BadT2I$_\textit{Sent}$ & VillanBKD$_\textit{one}$ & VillanBKD$_\textit{mul}$ &  PersonalBKD & EvilEdit   & \textbf{Avg}. & Iter./Sample \\
    \hline
    T2IShield$_\textit{FTT}$ 
    & \textbf{93.5}  & 49.2  & 50.8  & 46.7 & 71.9 &  80.6   & 47.6 & 49.5 & 61.2 & 50  \\
    T2IShield$_\textit{CDA}$ 
    & \blue{90.5}  & \blue{60.6}  & \blue{55.7}  & 48.7  & 78.9  & 84.1  & \blue{57.6}  & \blue{53.6}  & \blue{66.2}  & 50 \\
    UFID  
    & 58.7  & 46.7  & 48.4  & 51.1  & \blue{90.0}  & \textbf{98.0}  & 46.3  & 46.6  & 60.7 & 200 \\
    \rowcolor{gray!13}
    \shortNew  
    &  87.1  & \textbf{83.3}  & \textbf{89.9}  & \textbf{86.4}  & \textbf{97.1}  & \textbf{98.0}  & \textbf{93.5 } & \textbf{70.9}  & \textbf{88.3}  & $\approx$\textbf{10.3} \\
    \hline
    \end{tabular}%
    }
    \caption{The accuracy (ACC) of detection against the mainstream T2I backdoor attacks on Flickr~\cite{young2014image}. 
    The experiments are conducted on Stable Diffusion v1-5~\cite{Stable-Diffusion-v1-5}.
 We highlight key results as Tab.~\ref{tab:main}. 
 Since these values represent the ACC of a binary classification task, even "random guessing" achieves an ACC of 50.0\%.
 \textbf{Note that the threshold used here is the same as in \cref{tab:main_acc}}, 
 which demonstrates that the threshold computed in \cref{eq:th} based on our method can more effectively distinguish normal and poisoned samples across different data distributions.
}
  \label{tab:appd_main_acc}%
\end{table*}%

\noindent To further evaluate our method's performance across various data distributions and various models, we inject backdoors into the Stable Diffusion v1-5~\cite{Stable-Diffusion-v1-5} model and evaluate different detection methods using the Flickr~\cite{young2014image} dataset (\cref{tab:appd_main} and \cref{tab:appd_main_acc}). Note that since EvilEdit~\cite{wang2024eviledit} and PersonalBKD~\cite{huang2024personalization} can only target specific objects in the text, we continue to use the data as \cref{sec:exp} while replacing the model with Stable Diffusion v1-5~\cite{Stable-Diffusion-v1-5}. \textbf{Note that  we use the \underline{same detection threshold} as in \cref{tab:main_acc}, demonstrating the generalizability of our detection threshold.}

\vskip 0.2em
\noindent \textbf{Effectiveness Evaluation.}
In \cref{tab:appd_main} and \cref{tab:appd_main_acc}, it can be observed that,  our method also outperforms the baselines~\cite{wang2025t2ishield,guan2024ufid} similar to \cref{tab:main} and \cref{tab:appd_main_acc}, especially for the non-"entire image" backdoors in \cref{tab:appd_bkd}.
Compared to our method, the performance of the baselines on more stealthy backdoors is nearly equivalent to random guessing.

\vskip 0.2em
\noindent \textbf{Efficiency Evaluation.}
Similarly, to estimate computational overhead, we calculate the average non-stopword token length in the Flickr~\cite{young2014image} dataset, which is approximately $10.3$. This indicates that our method remains more efficient than the baselines, requiring only about 20\% time-cost of T2IShield~\cite{wang2025t2ishield} and 5\% time-cost of UFID~\cite{guan2024ufid}.

\section{Effects Under the More Advanced Adaptive Attack}\label{appd:further_adap}
Inspired by \cite{yi2024badacts}, we design a more advanced adaptive attack by adding a regularization term that enforces consistency constraints on activation variation. 
We implement the adaptive attack using $\text{BadT2I}_{\text{Tok}}$ and incorporate the regularization term from~\cref{eq:sum_layers}:
\begin{equation}
    \mathcal{L}_{\text{BadT2I\_Reg}} = \mathcal{L}_{\text{BadT2I}} +  \alpha \cdot \delta_{\theta} \left(c, c'\right),
\end{equation}
where $c$ and $c'$ denote the benign input and the trigger-embedded input, respectively. 
\begin{table}[h]
  \centering
      \footnotesize
\resizebox{0.9\linewidth}{!}{
\setlength{\tabcolsep}{13pt}
    \begin{tabular}{c|c|cccccc}
    \hline
    \textbf{\rule{0pt}{2.1ex}$\alpha$} & No Reg   & $10^{-5}$ & $10^{-7}$ & $10^{-8}$ & $10^{-9}$ & $10^{-10}$  \\
    \hline
    \textbf{FID} & 13.0      & 64.7 \red{(+397\%)}  & 16.8 \red{(+29\%)} & 15.0 \red{(+15\%)} & 13.2 & 13.1  \\
    \textbf{ASR} & 0.98       & 0.50  & 0.38  & 0.42  & 1.00  & 1.00 \\
    \textbf{FTR} & 0            & 0.42  & 0.24  & 0.13  & 0     & 0 \\
    \hline
    \textbf{AUC} & 97.0      & N/A     & N/A     & 81.2 & 93.6 & 95.7 \\
    \hline
    \end{tabular}%
    } 
      \caption{ The effectiveness of adaptive attacks against \shortNew under various weights.}
  \label{tab:reb_various_w}%
\end{table}%

Including the original weight $\alpha$ of 250 in~\cite{yi2024badacts}, 
we evaluate a set of  $\alpha$: 
$[250, 1, 10^{-3}, 10^{-5}, 10^{-7}, 10^{-8}, 10^{-9}, 10^{-10}]$.
We compute the \textbf{FID} of the backdoored model to assess its utility, \textbf{ASR} (Attack Success Rate) and \textbf{FTR} (False Triggering Rate) to assess backdoor effectiveness, and \textbf{AUC} of 
\shortNew. In Tab.~13, when $\alpha \geq 10^{-5}$, the model collapses and outputs noise. At $\alpha = 10^{-7}$ or $10^{-8}$, the model triggers the backdoor randomly, indicating unsuccessful backdoor injection. For $\alpha \leq 10^{-9}$, our method achieves satisfactory performance.
Hence, this adaptive attack is ineffective. 

\section{Comparison with Other Related Works}\label{appd:comparison}
\subsection{Backdoor Defenses for Unconditional Diffusion Models }\label{appd:comparison_uncondition}

In experiments (\cref{sec:effect}), we consider all existing T2I backdoor defense methods for comparison: T2IShield \cite{wang2025t2ishield} and UFID \cite{guan2024ufid}.  
There are also other backdoor defense methods~\cite{truong2025dual,mo2024terd,hao2024diff,wang2025lie} targeting diffusion models.  
However, we do not include them in our experiments because these methods are only applicable to unconditional diffusion models,  
and are not suitable for text-to-image synthesis scenarios (i.e., conditional diffusion models).
We select several works for discussion:  
\citet{truong2025dual,mo2024terd,hao2024diff} focus on inverting backdoor triggers in unconditional diffusion models (e.g., DDPM), where triggers are image distributions.  
In contrast, triggers in T2I diffusion models are textual tokens, making these methods inapplicable.  
\cite{wang2025lie} is also inapplicable as it aims to invert the visual trigger.


\subsection{Comparison with BadActs~\cite{yi2024badacts} }\label{appd:comparison_badact}

BadAct~\cite{yi2024badacts} is a similar work that utilizes neuron activations for backdoor defense on NLP classification models. Our work differs from it in  the following three aspects:
\ding{182} Methodologically, BadAct relies on {activation values} to detect outliers, whereas we apply token masking and calculate the {activation variations} to make judgments. 
\ding{183} Theoretically, BadAct does not involve the concept of diffusion.  
In contrast, we conduct a theoretical analysis to significantly improve detection efficiency.
\ding{184} Experimentally, BadAct is evaluated only on NLP models without T2I tasks.
So we evaluate it on T2I backdoors. In \cref{tab:reb_badact},
{\shortNew generally outperforms BadAct}. 
Specifically, BadAct produces predictions opposite to the gold-labels for $\text{BadT2I}_{\text{Tok}}$ inputs (similar observation in \cref{appd:nc2navi}). 
This is because the activation distribution of $\text{BadT2I}_{\text{Tok}}$ samples is more concentrated than that of clean inputs, causing BadAct to fail.
\begin{table}[t]
  \centering
      \resizebox{0.9\linewidth}{!}{
\setlength{\tabcolsep}{12pt}
\renewcommand{\arraystretch}{0.99}
    \begin{tabular}{c|c|cccc|c}
    \hline
   \rule{0pt}{2.05ex} & Backdoor & $\text{RickBKD}_{\text{TPA}}$ & $\text{BadT2I}_{\text{Tok}}$ & $\text{VillanBKD}_\textit{one}$ & EvilEdit & \textbf{Avg.} \\
    \hline
    \multirow{2}[2]{*}{AUC} & BadAct & 81.6  & \red{2.0}   & 97.4  & \red{62.0} & 60.7 \\
          & \shortNew & 99.9  & 97.0  & 98.9  & 85.5 & 95.3 \blue{(+34.6)} \\
    \hline
    \multirow{2}[2]{*}{ACC} & BadAct & 77.7  & \red{45.2}  & 87.3  & \red{52.0} & 65.6 \\
          & \shortNew & 91.2  & 91.4  & 94.5  & 71.7 &  87.2 \blue{(+21.6)} \\
    \hline
    \end{tabular}%
    } 
      \caption{The performance of BadAct against T2I backdoors.}
  \label{tab:reb_badact}%
\end{table}%

\section{Ablation Studies}\label{appendix:abla}

\subsection{From Neuron Coverage to Neuron Activation Variation}\label{appd:nc2navi}
Note that although we find that the NC value of trigger tokens differs from other tokens on an average scale, Neuron Coverage value~\cite{pei2017deepxplore} is too coarse-grained to be directly used for detecting backdoored samples. We design the following ablation experiment to validate this point.
\ding{182} We directly use the NC value of the input sample (a percentage value) as an indicator to determine whether it is a poisoned sample.  
\ding{183} We mask each token in the input sample and compute the maximum change in NC value as an indicator to determine whether it is a poisoned sample.
We report the AUROC value of detection in \cref{tab:nc2navi}.
We find that neither of the above methods achieves results as satisfactory as our approach, demonstrating that the layer-wise computation, ``neuron activation variation'', designed in \cref{sec:navi}  plays a crucial role.
\begin{table*}[ht]
  \centering
 \tablestyle{3pt}{1}
    \resizebox{\linewidth}{!}
  {
     \begin{tabular} 
     {@{}l|cccccccc|c@{}}
    \hline
    Method & RickBKD$_\textit{TPA}$ & RickBKD$_\textit{TAA}$ & BadT2I$_\textit{Tok}$ & BadT2I$_\textit{Sent}$ & VillanBKD$_\textit{one}$ & VillanBKD$_\textit{mul}$ & PersonalBKD   & EvilEdit  & \textbf{Avg}. \\
    \hline
    Neuron Coverage & 64.9  & 54.4  & 31.9  & 35.0  & 62.0  & 82.9  & 45.6  & 59.5  & 54.5 \\
    NC Variation & 91.5  & 71.6  & 64.8  & 71.5  & 84.4  & 56.5  & 65.8  & 62.8  & 71.1 \\ 
    \rowcolor{gray!13}
    \shortNew  & \textbf{99.9} & \textbf{99.8} & \textbf{ 97.0} & \textbf{89.7} & \textbf{98.9} & \textbf{99.9}  & \textbf{99.8} & \textbf{85.5}  & \textbf{96.3} \\
    \bottomrule
    \end{tabular}%
    }
     \caption{We compare the detection performance (AUROC) when using Neuron Coverage~\cite{pei2017deepxplore} instead of calculating ``Neuron Activation Variation'' (\cref{sec:navi}). Notably, the ``Neuron Coverage'' method even produces an opposite AUROC value against the BadT2I~\cite{zhai2023text} backdoors. This is because some backdoored samples increase the model's NC, while others decrease it.
}
  \label{tab:nc2navi}%
\end{table*}%

\subsection{Selection of Layers} 
\label{appd:layer_select}
\begin{table}[ht]
  \centering
  \resizebox{0.5\linewidth}{!}
  {
    \begin{tabular}{lcccc}
    \toprule
    Layer Selection &  VillanBKD$_\textit{mul}$ & BadT2I$_\textit{Tok}$ & EvilEdit & \textbf{Avg.} \\
    \midrule
    DownBlk & 99.9 & \cellcolor{red!20}  75.9 & \textbf{88.3} & 88.0\\
    
    MidBlk & 99.9 & 96.6 & \cellcolor{red!20} 76.3 & 90.9 \\
    
    UpBlk & 99.9 & 96.7 & 84.5 & 93.7 \\
    
    Attention-layers & 99.9 & 94.8 & 87.2 & 94.0 \\

    Conv-layers & 99.9 & \textbf{97.9} & 82.6 & 93.5 \\
    \rowcolor{gray!13}
    All layers & 99.9 & 97.0 & 85.5 & \textbf{94.1} \\
    
    \bottomrule
    \end{tabular}%
    }
      \caption{ Detection performance (AUROC) across different layer selections.}
  \label{tab:layer}%
\end{table}%
\noindent 
In the UNet model, the architecture is divided into three DownBlocks (DownBlk), one MidBlock (MidBlk), and three UpBlocks (UpBlk) based on 
resolution~\cite{Stable-Diffusion-v1-4, Stable-Diffusion-v1-5}, with each block containing attention layers, convolutional layers, and other linear layers. 
 We report ablation experiments for different layer selections within $\mathcal{L}_{set}$. 
 In \cref{tab:layer}, we find that our method exhibits reduced sensitivity to certain backdoors (in \textcolor[RGB]{255,78,78}{light red}) when using only the DownBlock and MidBlock. In contrast, utilizing all layers achieves the best overall performance. So we adopt this setting in \cref{eq:sum_layers}.

\subsection{Selection of Iteration Steps}\label{appd:step}
\begin{figure}[ht]
    \centering
    \includegraphics[scale=0.25]{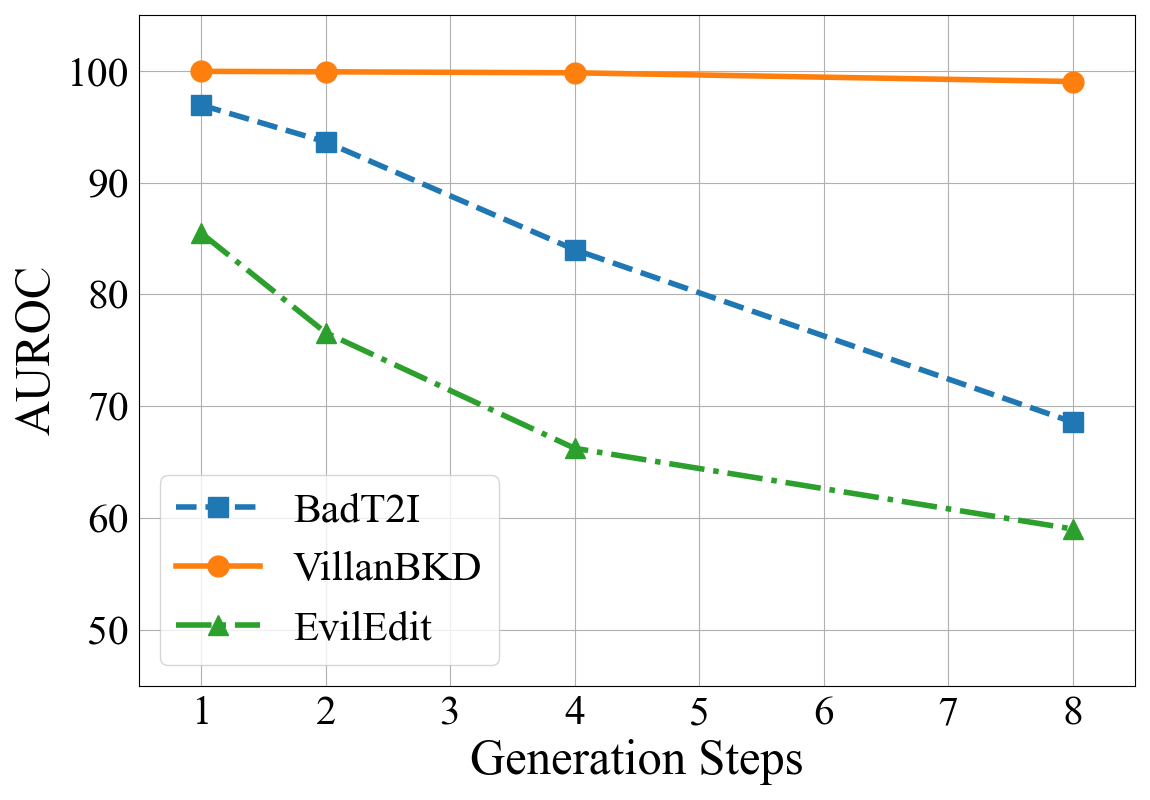}
\caption{Detection effectiveness at various generation steps. 
}
\label{fig:timestamps}
\end{figure}

 \noindent We evaluate the detection performance with different values of generating timestamps (\(T_{iter}=50\)), and report the AUROC values in Fig.~\ref{fig:timestamps}. We observe that as the timestamp increases, the AUROC value gradually decreases, which aligns with the \phenoName phenomenon in ~\cref{fig:intuition}. 

\subsection{Hyperparameter for Score Function in \cref{eq:div_percent}}\label{appd:hyper_score}
In \cref{eq:div_percent}, we exclude elements above the 75th percentile and compute the average value. This approach aims to eliminate outlier values, which may be related to trigger tokens, while the remaining elements are more likely to represent normal token values. Here, we analyze the impact of different percentile choices on the detection results.
We conduct additional ablation experiments to analyze the impact of different percentile parameters:  
\ding{182} ``Mean'' – We directly compute the mean without excluding outliers.  
\ding{183} ``ExMax'' – We compute the mean of all elements except the maximum value.  
\ding{184} ``75th'' - We exclude elements above the 75th percentile and compute the mean.  
\ding{185} ``50th'' - We exclude elements above the 50th percentile and compute the mean.
We report the AUROC values of different methods in \cref{tab:appd_percent}.
The experimental results show that different parameter choices have only a minor impact on the performance, which demonstrates the robustness of our method to
hyperparameter selection. We adopt the 75th percentile in \cref{eq:div_percent} because it exhibits a slight advantage over other parameter choices.

\begin{table*}[ht]
  \centering
 \tablestyle{3pt}{1}
    \resizebox{\linewidth}{!}
  {
     \begin{tabular} 
     {@{}l|cccccccc|c@{}}
    \hline
    Percentile & RickBKD$_\textit{TPA}$ & RickBKD$_\textit{TAA}$ & BadT2I$_\textit{Tok}$ & BadT2I$_\textit{Sent}$ & VillanBKD$_\textit{one}$ & VillanBKD$_\textit{mul}$ & PersonalBKD   & EvilEdit  & \textbf{Avg}. \\
    \hline
   
    Mean  & 99.7  & 99.7  & 96.8  & 88.7  & 98.8  & 99.9  & 99.8  & 84.8  & 96.02 \\
    ExMax & 100.0 & 99.8  & 97.3  & 88.8  & 98.8  & 99.9  & 99.9  & 84.8  & 96.15 \\
    \rowcolor{gray!13} 75th  & 99.9  & 99.8  & 97.0  & 89.7  & 98.9  & 99.9  & 99.8  & 85.5  & \textbf{96.31} \\
    50th  & 99.0  & 99.9  & 96.6  & 90.2  & 99.1  & 99.9  & 99.9  & 85.3  & 96.24 \\

    \bottomrule
    \end{tabular}%
    }
     \caption{The performance (AUROC) of different percentile choices in \cref{eq:div_percent}.
     The experiments demonstrate the robustness of our method to hyperparameter selection.
}
  \label{tab:appd_percent}%
\end{table*}%

\subsection{Hyperparameter for Threshold in \cref{eq:th}}\label{appd:hyper_th}

\begin{table*}[ht]
  \centering
 \tablestyle{3pt}{1}
    \resizebox{\linewidth}{!}
  {
     \begin{tabular} 
     {@{}l|cccccccc|c@{}}
    \hline
    Value of m in \cref{eq:th}  & RickBKD$_\textit{TPA}$ & RickBKD$_\textit{TAA}$ & BadT2I$_\textit{Tok}$ & BadT2I$_\textit{Sent}$ & VillanBKD$_\textit{one}$ & VillanBKD$_\textit{mul}$ & PersonalBKD   & EvilEdit  & \textbf{Avg}. \\
    \hline

    m=1.1 & 89.8  & 90.5  & 90.9  & 80.4  & 95.2  & 98.3  & 95.6  & 73.0  & 89.21 \\
    \rowcolor{gray!13} m=1.2 & 91.2  & 91.8  & 91.4  & 79.2  & 94.5  & 98.9  & 95.6  & 71.7  & \textbf{89.29} \\
    m=1.3 & 92.3  & 92.7  & 90.7  & 78.3  & 94.1  & 99.0  & 93.5  & 70.6  & 88.90 \\
    m=1.5 & 93.5  & 94.8  & 90..4 & 76.0  & 93.3  & 99.0  & 92.8  & 66.8  & 77.03 \\

    \bottomrule
    \end{tabular}%
    }
     \caption{The performance (ACC) of different values of $m$ in \cref{eq:th}.
}
  \label{tab:appd_m_acc}%
\end{table*}%

\noindent We select different values for \( m \) in \cref{eq:th} and report the detection performance (ACC) of our method in \cref{tab:appd_m_acc}. Since \cref{eq:th} essentially characterizes the boundary for outlier data within clean samples, a larger \( m \) tends to classify more samples as clean data, while a smaller \( m \) tends to classify more samples as poisoned data. According to the results in \cref{tab:appd_m_acc}, the optimal classification performance is achieved when \( m \) is set to 1.2.

\section{Another view of \cref{theorem:diffusion_prediction_similar}}\label{appd:another_view}

In this section, we provide another view of \cref{theorem:diffusion_prediction_similar}. In \cref{theorem:diffusion_prediction_similar}, we bound the error \(\epsilon\) by some function of \(\alpha\) and \(\sigma_t\). In this section, we would provide another dual theorem, which shows that for any error \(\epsilon\), there always exists a critical point \(t^*\), such that for any \(t<t^*\), the prediction of diffusion model under different conditions is \(\epsilon\)-similar.

\begin{theorem}
   Assume the diffusion model is well-trained, i.e., achieving the minimal \(\mathbb{E}[\|\epsilon(\vect{x}_t,t,\mathbf{c})-\epsilon\|_2]\) on some discrete distribution. As long as \(p(\mathbf{c}|\vect{x})\) is not strictly 1 or 0, i.e., there exists \(\alpha>0\) such that \(\alpha\leq p(\vect{x}|\mathbf{c})\leq1-\alpha\) for any input \(\vect{x}\), two different condition \(\mathbf{c}, \mathbf{c}'\), then, for any error \(\epsilon\), there always exists a critical point \(t^*\), such that for any \(t<t^*\), the prediction of diffusion model under different condition are \(\epsilon\)-similar:
   \begin{equation*}
       \|\epsilon(\vect{x}_t,t,c)-\epsilon(\vect{x}_t,t,c')\|_2 \leq \epsilon,
   \end{equation*}
    where
    \(\sigma_{t^*}=O(\sqrt{\frac{1}{-\ln\alpha \epsilon}})\) and $c'$ is another condition embedding from text $p'$: $c' = \mathcal{T}(p')$.
\label{theorem:diffusion_prediction_similar_t}
\end{theorem}

Note that \(\frac{1}{-\ln\alpha \epsilon}\) is a much slower decay rate than any polynomial rate \(\frac{1}{\text{poly}(\epsilon, \alpha)}\). This indicates that the diffusion model's different predictions would quickly become extremely similar. This is aligned with the empirical observation that the cosine similarity between the prediction of diffusion models becomes more than \(0.9999\) even after just 8 sampling steps.

\begin{lemma}
    There always exists a target image \(\vect{x}_{\text{final}} \in \mathcal{D}\), such that for any error \(\epsilon_s\), there always exists a critical point \(t^*\), such that for any \(t<t^*\), we have \(s_{\mathcal{D},c}(\vect{x}_{\text{final}}) > 1-\epsilon_s\).
\label{appendix:lemma:2}
\end{lemma}

This lemma indicates that, when the sampling process proceeds, the posterior distribution gradually becomes a Dirac distribution, converging to one point in the training set.

\begin{proof}
    Let \(\vect{x}\) be the closed point in dataset from \(\vect{x}_t\), i.e., \(\min_{\vect{x} \in \mathcal{D}} \|\vect{x}-\vect{x}_t\|^2\), and \(\vect{x}_2\) be the second closest point.
    \begin{align*}
        &s_{\mathcal{D},c}(\vect{x}) \geq 1-\epsilon_s \Leftrightarrow \frac{\exp(-\frac{\|\vect{x}_t-x\|^2}{2\sigma_t^2}+\log q(\vect{x}|c))}{\sum_{\vect{x}'} \exp(-\frac{\|\vect{x}_t-\vect{x}'\|^2}{2\sigma_t^2}+\log q(\vect{x}'|c))} \geq 1-\epsilon_s \\
        \Leftrightarrow& \frac{1}{1+\sum_{\vect{x}' \neq \vect{x}} \exp(\frac{\|\vect{x}_t-\vect{x}\|^2}{2\sigma_t^2}-\frac{\|\vect{x}_t-\vect{x}'\|^2}{2\sigma_t^2}+\log q(\vect{x}'|c)-\log q(\vect{x}|c))} \geq 1- \epsilon_s \\
        \Leftrightarrow& \sum_{\vect{x}' \neq \vect{x}} \exp(\frac{\|\vect{x}_t-\vect{x}\|^2}{2\sigma_t^2}-\frac{\|\vect{x}_t-\vect{x}'\|^2}{2\sigma_t^2}+\log q(\vect{x}'|c)-\log q(\vect{x}|c)) \leq \frac{\epsilon_s}{1-\epsilon_s}.
    \end{align*}
    This can be relaxed to:
    \begin{equation*}
        (|\mathcal{D}|-1) \exp(\frac{\|\vect{x}_t-\vect{x}\|^2}{2\sigma_t^2}-\frac{\|\vect{x}_t-\vect{x}_2\|^2}{2\sigma_t^2}+\log q(\vect{x}_2|c)-\log q(\vect{x}|c)) \leq \frac{\epsilon_s}{1-\epsilon_s}.
    \end{equation*}
    As long as \(p(c|\vect{x})\) is not strictly 1 or 0, i.e., there exists an \(\alpha\) such that \(0< \alpha \leq p(\vect{x}|c) \leq 1-\alpha < 1\), we can further relax to:
    \begin{align*}
        & (|\mathcal{D}|-1) \exp(\frac{\|\vect{x}_t-\vect{x}\|^2}{2\sigma_t^2}-\frac{\|\vect{x}_t-\vect{x}_2\|^2}{2\sigma_t^2}+\log (1-\alpha)-\log \alpha \leq \frac{\epsilon_s}{1-\epsilon_s} \\
        \Leftrightarrow& \frac{\|\vect{x}_t-\vect{x}\|^2-\|\vect{x}_t-\vect{x}_2\|^2}{2\sigma_t^2} \leq \log \frac{\epsilon_s}{(D-1)(1-\epsilon)}+\log \frac{\alpha}{(1-\alpha)}=O(\log \alpha \epsilon_s).
    \end{align*}
    Therefore, if we want \(s_{\mathcal{D},c}(\vect{x}_{\text{final}}) > 1-\epsilon_s\), we can get the condition for \(\sigma_t^2\):
    \begin{align*}
        \sigma_t^2 \leq O(\frac{1}{\log \alpha \epsilon_s}) \Leftrightarrow \sigma_t \leq O(\sqrt{\frac{1}{\log \alpha \epsilon_s}}).
    \end{align*}
\end{proof}

Therefore, by Lemma \ref{appendix:lemma:epsilon_s_and_epsilon}, to let \(\|h(\vect{x}_t,t,c)-h(\vect{x}_t,t,c')\|_2^2 \leq \epsilon\), we require \(3 \epsilon_s C^2\leq \epsilon\), that is \(\epsilon_s \leq \frac{\epsilon}{3C^2}\). We can get the requirement for \(\sigma_t\):
\begin{equation*}
    \sigma_t \leq O(\sqrt{\frac{1}{\log \alpha \frac{\epsilon}{3C^2}}}) = O(\sqrt{\frac{1}{\log \alpha\epsilon}}).
\end{equation*}


\end{document}